\documentclass{ws-rv9x6}
\usepackage{ws-rv-van,bm}             
\makeindex

\begin{document}

\chapter[Coherent control in QDs]{Coherent control and decoherence of
charge states in quantum dots}

\author[P. Machnikowski]{Pawe{\l} Machnikowski}

\address{Institute of Physics, Wroc{\l}aw University of Technology,\\
50-370 Wroc{\l}aw, Poland, \\
Pawel.Machnikowski@pwr.wroc.pl}

\begin{abstract}
This Chapter contains a review of the recent results, both
experimental and 
theoretical, related to optical control of carriers confined in
semiconductor quantum dots. The physics of Rabi
oscillations of exciton and biexciton occupations, as well as 
time-domain
interference experiments are discussed. Next, the impact of carrier--phonon
interaction in a semiconductor structure is described and modern
methods of theoretical description of the carrier--phonon kinetics and
of the resulting dephasing are presented.
\end{abstract}

\body

\section{Introduction}\label{sec:intro}

The progress of semiconductor technology that took place in the 80s
and 90s of the last century, in particular the rapid
development of epitaxy and lithography techniques, has allowed
physicists to manufacture and study structures in
which carriers are confined to a small volume in space (tens of
nanometers or even less) \cite{reed93}.
In this way quantum dots (QDs)\index{quantum dot}, 
that is, artificial structures (boxes)
containing a few
particles with quantized energy levels, have been produced. Because of
the similarity to natural atoms, such structures are also referred to
as artificial atoms. However, the properties of QDs 
are more flexible in comparison with atoms: 
their shapes, size
and various other features can be engineered at the stage of
manufacturing by modifying the technological conditions and the number
of confined electrons can be changed under laboratory conditions
(e.g. by applying an external voltage) within a wide range of values.

Parallel to the laboratory investigations, a rapid process 
of miniaturization of commercial semiconductor structures
(e.g., computer chips) took place.
At this moment, the state-of-the-art commercially available microchips,
manufactured using the immersion photolithography technique, are
built of elements whose size can be reduced down to the 90 nm
diffraction limit. The introduction of the 45 nm process is announced
for 2007 or early 2008 and the implementation of the 16 nm technology is
envisaged in the time frame 2013-14.\footnote{International
Technology Roadmap for Semiconductors (www.itrs.net).} Thus, the
characteristic size of elements in the microchips of our standard
computer equipment has dropped below the size of the first QDs
obtained in laboratories 20 years ago and rapidly approaches the size of the
smallest structures described in today's research papers.

This progress of manufacturing technology is accompanied by a rapid
development of optical spectroscopy. Currently, it is
possible to study the optical properties of a single QD and to
coherently control the evolution of a single carrier or a pair of
carriers (electron-hole pair) in such a
structure. Many experimental schemes of quantum optics have been
implemented on QDs. Moreover,
certain procedures relying on the specific structure of the energy
levels of these artificial semiconductor structures have been
demonstrated, which have no counterpart in atomic systems. These
experimental achievements have motivated theoretical proposals for
sophisticated quantum-optical schemes that may lead, for instance, to
optical control of a single electron spin in a QD.

The goal of this chapter is twofold: First, to introduce the reader
into the fascinating world of modern optical experiments on
semiconductor quantum dots and to the astonishingly simple, yet
nontrivial, theory underlying the phenomena observed in the
labs on the fundamental, quantum-optical level. 
Second, to give a review of some more sophisticated theoretical methods
that allow one to include the interaction with the lattice vibration
modes (phonons) which are specific to semiconductor systems.

\section{Essential properties of quantum dots}\label{sec:qd}

Quantum dots are semiconductor structures in which the carrier
dynamics is restricted in all three dimensions to the length scales of
several or a few tens of nanometers \cite{jacak98a,bimberg99}. 
Various structures that have this property may be obtained by a
variety of methods. 

\begin{figure}[tb]
\centerline{\psfig{file=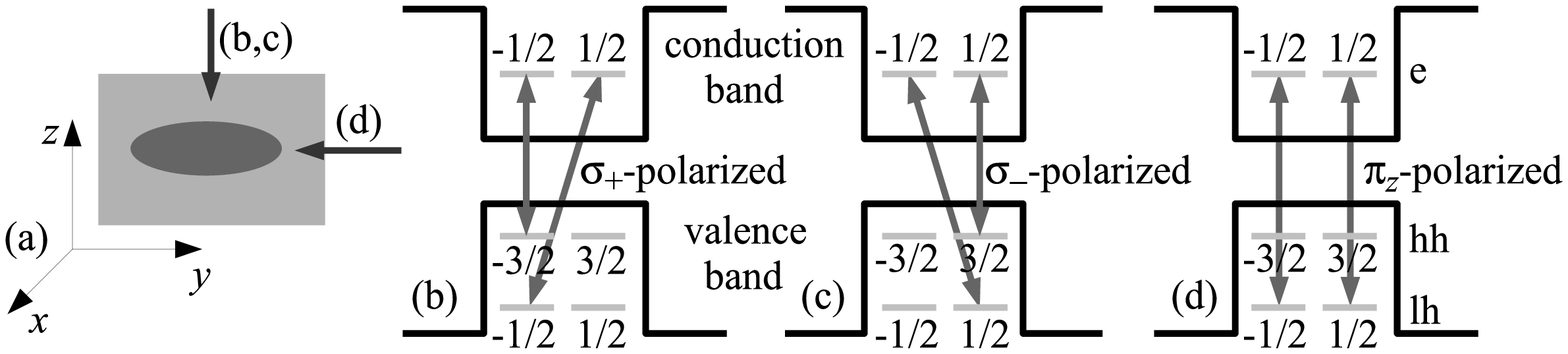,width=11cm}}
\caption{(a) Schematic plot of a QD. The $xy$ plane is referred to as
the structure plane, while z is called growth direction and
corresponds to the (approximate) symmetry axis of the structure. The
arrows show the 
incidence of light corresponding to the three diagrams b-d. (b-d) The
diagrams showing the transitions allowed by selection rules in a III-V
QD, induced by light circularly polarized in the structure plane (b:
right-polarized, c: left-polarized) and by light linearly polarized
along z (d).} 
\label{fig:sel-rules}
\end{figure}

One of the most widely used techniques is the
Stransky-Krastanov self-assembly. 
When a semiconductor compound is epitaxially
(layer by layer) grown on a substrate with a different lattice
constant (the InAs/GaAs pair is a typical example) each new layer must
be squeezed to match the lattice constant of the substrate. At some
point (at about 1.7 monolayers for InAs/GaAs) it is energetically favorable for the
epitaxial layers to restructurize into a system of islands, which
increases the free surface but relaxes strain  \cite{petroff94}. The
sample is then 
covered with the substrate material, leading to lens-shaped (or,
sometimes, pyramidal) nanostructures as in 
Fig.~\ref{fig:sel-rules}a \cite{fry00}.
In general, the band edges of the nanostructure material are offset with
respect to those of the substrate. Here we will only discuss
structures as those in  Fig.~\ref{fig:sel-rules}b-d, where the
conduction band edge is shifted down and the valence band edge is
shifted up, so that both electrons and holes are bound in the QD structure.

Another type of structures commonly used in the optical experiments
are thickness fluctuations of a thin epitaxial layer of a
semiconductor (so called quantum well) placed between thick structures
of different semiconductor with a wider band gap. Here, again, the band
edge offset leads to localization of carriers within the quantum well
layer. If the epitaxial growth of the quantum well layer has been
stopped after the formation of a new monolayer started, the quantum
well has one-monolayer thickness fluctuations which weakly localize
the carriers. 

The QD nanostructures are typically 2--3 orders of magnitude larger
than atoms. However, the effective mass of
carriers in a semiconductor is often considerably lower than the
free electron mass (e.g. $m^{*}=0.07m_{0}$ in GaAs), and
this degree of confinement is sufficient for quantization of carrier
energies with electron inter-level spacing reaching 100 meV in
self-assembled structures. This is definitely enough to resolve the
states spectrally and to neglect thermal transitions to the excited
states even at moderate temperatures.
Therefore, we will restrict the discussion to the ground state of each
kind of carriers. 

The exact properties of the quantum states in a QD, e.g., the geometry of
wave functions or Coulomb interaction between the confined carriers,
may be found, e.g., by tight-binding or pseudopotential calculations.
For lens-shaped QDs, a simple 2-dimensional harmonic model
\cite{jacak98a} has been shown to be a very good approximation
\cite{wojs96}. The states of an interacting few-particle system may then
be found by numerical configuration--interaction techniques
\cite{wojs95,hawrylak96}.
However, these details are irrelevant for
the discussion on the general quantum-optical level. 
Whenever a specific model is necessary
(Secs.~\ref{sec:phonons} and \ref{sec:kinet}) we will use simple
Gaussian wave functions. Also in these cases, the results do not
depend essentially on this choice.

Even with the restriction to the lowest orbital states, the structure
of energy levels and allowed optical transitions in a QD becomes quite
complicated if the angular momenta of carriers are taken into account.
The valence band in III-V semiconductors
is composed of p-type atomic orbitals (orbital angular momentum $1$),
yielding six quantum states (taking spin into account) for each
quasi-momentum $\bm{k}$. Due to considerable 
spin-orbit coupling\index{spin-orbit coupling} the
orbital angular momentum and spin are not separate good quantum
numbers and the valence band states of a bulk crystal must be
classified by the total angular momentum and its projection on a
selected axis.  Thus, the valence band is composed of three sub-bands
corresponding to two different representations of the total angular
momentum $J$. Out of these, the two states with $j=1/2$ form a subband
which is split-off by the spin--orbit interaction. The other four
states with $j = 3/2$ are degenerate in a bulk crystal at $\bm{k} = 0$ but this
degeneracy is lifted by size quantization and strain in a QD
structure, with the heavy hole (hh)\index{hole}\index{heavy hole} 
subband (angular momentum
projection on the symmetry axis ($m = \pm 3/2$) lying above the light hole
(lh)\index{hole}\index{light hole} 
subband ($m = \pm 1/2$) in all known structures.  The fundamental
optical excitation of a QD consists in transferring optically an
electron from the highest confined state in the valence band (thus
leaving a hole) to the lowest confined state in the conduction
band. The interacting electron--hole pair created in this way is
referred to as exciton\index{exciton} 
(lh-exciton or hh-exciton, depending on the
kind of a hole involved). Angular momentum selection
rules\index{selection rules}
restrict the
transitions allowed for a given propagation direction and polarization
of the light beam, as depicted in Fig.~\ref{fig:sel-rules}. 
For instance, according to
the selection rules represented in Fig.~\ref{fig:sel-rules}b, 
a right circularly polarized ($\sigma_{+}$-polarized) laser beam
can only create an exciton\index{exciton} with total momentum +1, referred to as 
``$\sigma_{+}$ exciton'', in accordance with the angular momentum conservation
(removing an electron with the angular momentum m is equivalent to
the creation of a hole with the angular momentum $-m$). Similarly, a
$\sigma_{-}$-polarized (left circularly polarized) 
beam creates only a ``$\sigma_{-}$ exciton''
with the angular momentum $-1$. Although there are still two
transitions allowed for a given circular polarization they can easily be
distinguished since the lh states are well separated
energetically from the lowest hh states.

In appropriately doped structures, QDs in the ground
state of the system may be occupied by electrons. An optical excitation in
this case corresponds to a transition between a single electron state
and a negative trion\index{trion} state, i.e., the state of two electrons and one
hole confined in a QD. From the Pauli exclusion principle it is clear
that this transition is possible only if the state which is to be
occupied by the photo-created electron is free. Hence, in the situation
of Fig. 1b, a heavy hole trion\index{trion} may be created if the dot is initially
occupied by a ``spin up'' ($m=+1/2$) electron but not if the electron in the
dot is in the ``spin down'' ($m=-1/2$) state. This suppression of the
optical transition depending on the spin of the electron in the QD is
referred to as Pauli blocking\index{Pauli blocking} 
and has been indeed observed
experimentally \cite{chen00}.

\begin{figure}[tb]
\centerline{\psfig{file=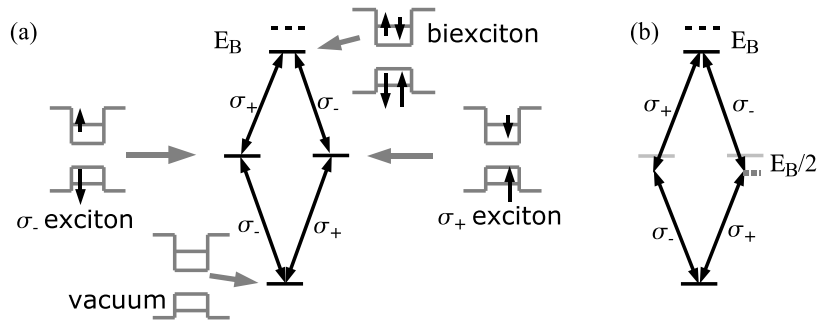,width=11cm}}
\caption{(a) The four states of the heavy-hole biexciton system and
the optical transitions between them. The band diagrams show the
particles forming each state. The arrows in the valence band represent
holes. (b) Schematic representation of the
two-photon resonance between the ground state and the biexciton state,
achieved with a properly tuned linearly polarized laser beam. Note that
the single exciton transitions are forbidden.} 
\label{fig:biex-diagram}
\end{figure}

After exciting an electron from the heavy hole band with a
$\sigma_{+}$-polarized laser beam ($m=-3/2\to -1/2$), as in
Fig.~\ref{fig:sel-rules}b,  it is still possible to transfer also
the $m=+3/2$ electron to the $m=+1/2$ conduction band state using
$\sigma_{-}$-polarized light (Fig.~\ref{fig:sel-rules}c). 
Thus, if the optical processes are spectrally
restricted to heavy holes (as is the case in most experiments) there are
four optically active states linked by allowed optical transitions as
shown in Fig.~\ref{fig:biex-diagram}a.  The highest state, with two
electrons 
and two holes present in the QD, is treated as composed of two
electron-hole pairs and is called a biexciton\index{biexciton}.
The Coulomb (dipole-dipole) interaction between the two
excitons\index{exciton} shifts the energy of the biexciton\index{biexciton} 
state\cite{hartmann00a} by the binding energy $E_{\mathrm{B}}$, 
so that the exciton\index{exciton}--biexciton transitions
are non-degenerate with the ground state--exciton\index{exciton} transitions. In this
way, all four transitions in the diagram can be distinguished either
spectrally or by polarizations.

\section{Coherent control: experimental state of the
art}\label{sec:exp}

The recent progress in ultrafast spectroscopy of semiconductor
systems \cite{axt04,krenner05} made it possible to control and probe the
quantum states of carriers confined in a QD on femtosecond time
scales. In particular, 
high degree of control over carrier occupations
in various kinds of QD structures has been demonstrated.
One kind of an experiment consists in measuring the average
occupation of the QD after a pulse of fixed length but variable
amplitude. 
The QD occupation is defined as the probability of
finding an exciton\index{exciton} in the QD after the laser pulse, calculated as 
the fraction of cases in which an exciton\index{exciton} was created over a large number of
repetitions of the experiment.
Within the linear absorption theory, the excitation (the QD
occupation) 
grows proportionally to the pulse intensity. Obviously, this
growth cannot be unlimited. When the occupation of the excited level
is sufficiently large the spontaneous and induced emission processes
suppress further increase of the occupation. Here, we are interested in
the coherent limit, where the evolution is induced by a strong
(essentially classical) laser field, inducing large occupation changes
over time scales much shorter than the spontaneous emission
time. Then, the system is driven from the ground to the excited state
in a coherent way, and than back to the ground state via a coherently
induced emission process. As a result, in an ideal case, the
final occupation after a pulse should show sinusoidal oscillations
between 0 and 1 as a function of the square root of pulse intensity
(referred to as \textit{pulse area} -- see
Sec.\ref{sec:two-level-rabi}). Such oscillations, known as pulse area dependent
Rabi oscillations\index{Rabi oscillations}, 
have been indeed observed in a range of experiments
on single QDs
\cite{stievater01,htoon02,zrenner02,stufler05a}. Similar
effect can also be observed in ensembles of QDs \cite{borri02a}.

\begin{figure}[tb]
\begin{center}
\parbox{6cm}{\epsfig{figure=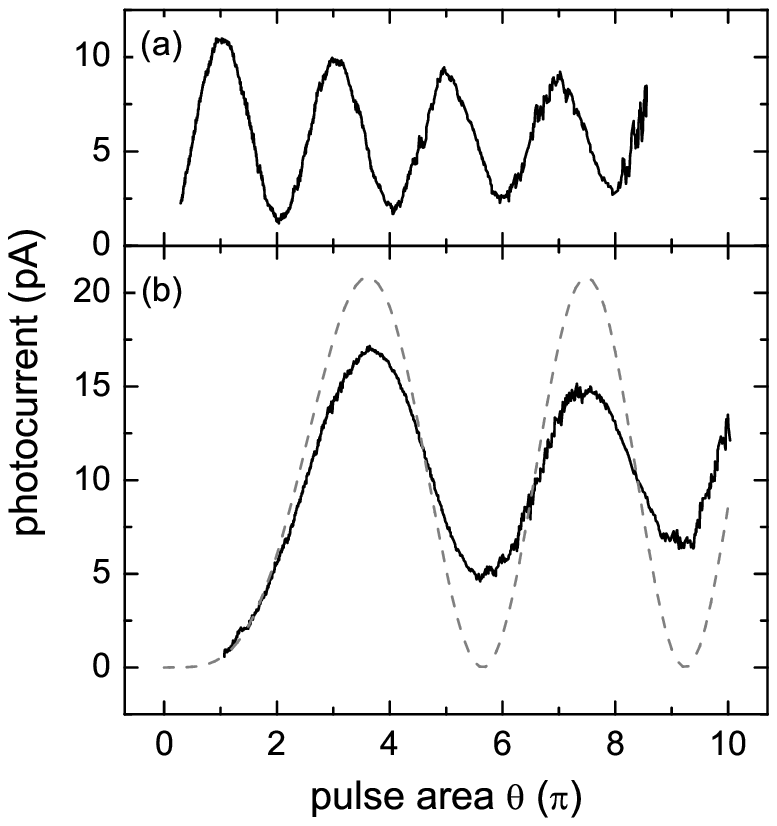,width=6cm}}
  \hspace*{1.5cm}
  \parbox{3.2cm}{\epsfig{figure=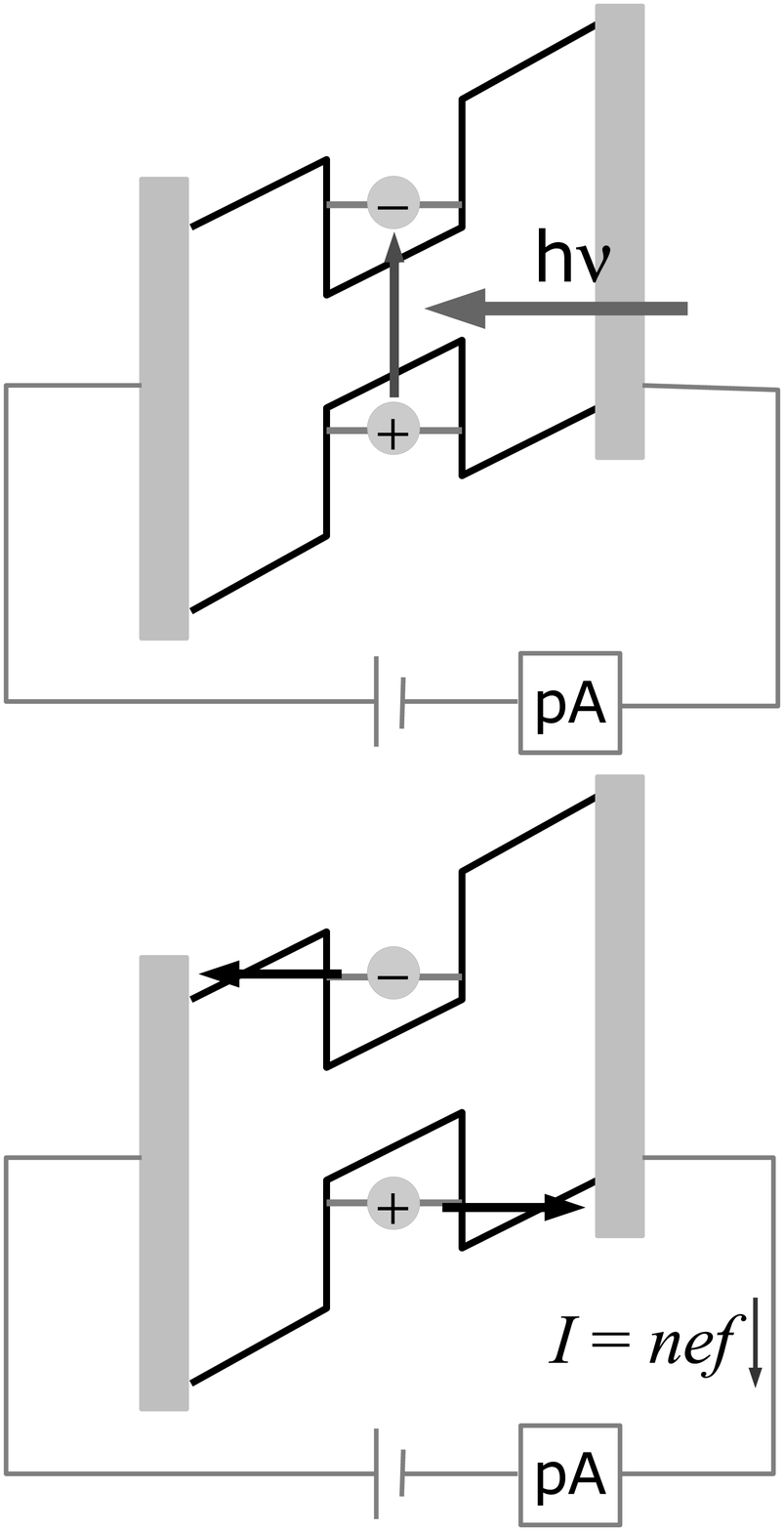,width=3.2cm}
  \figsubcap{c}}
\end{center}
\caption{
(a) Pulse area dependent Rabi oscillations. (b) Two-photon Rabi
oscillations. Solid line: experiment, dashed line: theory (without
dephasing), assuming a sech pulse shape with a pulse length of 2.3 ps and
biexciton binding energy $E_{\mathrm{B}}=2.7$~meV. Reprinted from 
Ref.~\refcite{stufler06}, S. Stufler \textit{et al.},
Phys. Rev. B {\bf 73}, 125304  (2006), \copyright American Physical
Society 2006. (c) A schematic explanation of
the functionality of the QD photodiode.}
\label{fig:Rabi}
\end{figure}

In Fig.~\ref{fig:Rabi}a we show the results of such an experiment,
performed with a single QD-based photodiode\index{photodiode} structure 
\cite{zrenner02,stufler05a,stufler06}. In this experiment the QD is
placed in an electric field between a pair of electrodes 
(Fig.~\ref{fig:Rabi}c). Each
electron-hole pair generated by the optical excitation contributes to
the photocurrent in the structure. If the laser pulse repetition rate
is $f$ (typically in kHz--MHz range) and the average QD occupation for
a given pulse area is $n$, then the repeated pulsing results in a
current $I=nef$. In this way, the Rabi oscillations\index{Rabi oscillations}
of the average exciton\index{exciton} number are reflected in the oscillating form of
the photocurrent as a function of the pulse area, which provides a
means of detection much more efficient than optical ones. These
oscillations are clearly seen in Fig.~\ref{fig:Rabi}a, although some
damping, due to dephasing, is also visible.

Apart from the coherent control of exciton\index{exciton} occupations, 
it has also been shown that phase control of carrier states in QDs is
possible. A laser pulse detuned from the exciton\index{exciton} resonance cannot
induce real transitions and, therefore, does not change
occupations. Nonetheless, as shown in an optical experiment with
interface fluctuation QDs \cite{unold04}, it can shift the energy
levels via the AC 
(optical) Stark effect and affect the phases in a quantum superposition of
empty dot and single-exciton\index{exciton} states. 
Another way to control the phases is to drive the system with a 
slightly detuned laser pulse, which leads to a combination of
occupation and phase evolution \cite{vasconcellos06} (see
Sec.~\ref{sec:two-level-rabi}). In this case,
phase effects can be observed in the form of Ramsey
interference fringes\index{Ramsey fringes} \cite{stufler06b}. 

It is possible to see in an experiment that the quantum state of an
electron-hole pair in a QD maintains its phase coherence long after
the laser pulse has been switched off. To this end, one splits the
pulse into two parts. If the phase of the quantum state in the QD were
random when the second pulse arrives the phase of the latter would be
irrelevant.
On the contrary, in the experiment one observes oscillations of the
final QD occupation as a function of the relative phase shift between the
two pulses \cite{bonadeo98,kamada01,htoon02} 
(the phase is shifted by tuning the delay between the
pulses with a sub-femtosecond accuracy, that is, by a fraction of the
optical oscillation period). Such an effect is referred to as
time-domain interference\index{time-domain interference}, 
since it may be interpreted in terms of
interference between the probability amplitudes for exciting the QD
with the first or with the second pulse. A formal treatment of this
class of experiments will be given in Sec.~\ref{sec:2level-interf}.

Apart from the fact that such interference experiments demonstrate coherent
phase-sensitive quantum control with an amazingly precise timing, they
are also of interest from a more general point of view. 
Obviously, in spite of the
wave-like behavior manifested by the interference effect, 
a single measurement of the QD occupation always yields either 0 or
1, demonstrating the particle-like nature of the exciton\index{exciton}.
Thus, the time-domain interference
experiments on QDs demonstrate quantum complementarity
between the particle-like nature of an exciton\index{exciton} and 
its ability to show quantum interference \cite{machnikowski06c}.

Controlling a single quantum degree of freedom (an exciton\index{exciton}) is just the
first step towards large-scale nano-optoelectronic and quantum
computing applications.  The next step towards more complex
implementations is to include coupling between two or more individual
quantum subsystems. The simplest experimental realization of such a
composite system is a biexciton\index{biexciton}. As explained in 
Sec.~\ref{sec:qd},
due to the Coulomb interaction between the two excitons\index{exciton}, the
excitation energy of an exciton\index{exciton} in the presence of the other one is
different than in its absence. Thus, these two transitions can be
addressed individually. Together with the polarization dependence of
the allowed transitions (selection rules, see Sec.~\ref{sec:qd}), this
allows one to excite, say, the $\sigma_{+}$ exciton if and only if the
the $\sigma_{-}$ exciton\index{exciton} is present. 
Such a conditional control procedure was indeed performed in an
experiment \cite{li03}, where Rabi oscillations\index{Rabi oscillations} on the
exciton\index{exciton}--biexciton\index{biexciton} 
transition were demonstrated.
If the two excitons\index{exciton} are viewed as
quantum bits\index{qubit} 
(with 0 and 1 corresponding to their presence or absence
in the QD) then such a conditional excitation constitutes an
implementation of the controlled-NOT gate\index{controlled-NOT gate}
which is fundamental for
quantum commuting \cite{nielsen00}. Such a controlled-NOT gate was indeed
implemented experimentally in a QD system \cite{li03}.
In a similar way it is possible to coherently manipulate a
biexciton system in two coupled QDs (with one exciton\index{exciton} localized in
each dot) \cite{unold05}.

It is interesting to see what happens if the frequency $\omega$ of a
linearly polarized laser beam is chosen such
that the energy of two photons matches the biexciton\index{biexciton} energy,
$2\omega=2E-E_{\mathrm{B}}$, 
while the single-exciton\index{exciton} transitions are detuned by $E_{\mathrm{B}}/2$
(Fig.\ref{fig:biex-diagram}b). A linearly polarized pulse is a
superposition of two circularly polarized components, so that both
exciton\index{exciton} transitions are allowed by the selection rules but neither of
them satisfies the energy conservation. Perturbation theory would
predict that the occupation of the biexciton\index{biexciton} state should grow
proportionally to the square of the pulse intensity (that is, to the
4th power of the pulse area), as a result of a two-photon absorption
process. Experimental results \cite{stufler06} presented in
Fig.~\ref{fig:Rabi}b indeed show
such a behavior for low pulse intensities. However, beyond this
perturbative regime, a pattern of two-photon pulse area dependent Rabi
oscillations between the ground and biexciton\index{biexciton} states develops, which
become almost periodic for large pulse intensities. The theory of such
coherent phenomena in the biexciton\index{biexciton} system will be presented in
Sec.~\ref{sec:biex}. 

\section{Quantum dot as a two-level system}
\label{sec:two-level}

Let us now proceed to a theoretical description of the quantum
evolution of carriers confined in a QD and subject to a laser
field. We will start with the simplest situation when the QD is driven
by a circularly polarized laser beam (say, $\sigma_{+}$) tuned to the
fundamental heavy hole transition, as in Fig.~\ref{fig:sel-rules}a
(vertical arrow). Then, only two states are involved in the evolution:
the ground state (empty dot), denoted by $|0\rangle$,
and the $\sigma_{+}$ exciton\index{exciton} state, denoted $|1\rangle$ (see
also Fig. \ref{fig:biex-diagram}a). Therefore, one effectively deals
with a very simple two-level system.

\subsection{General considerations}
\label{sec:2level-general}

\begin{figure}[tb]
\centerline{\psfig{file=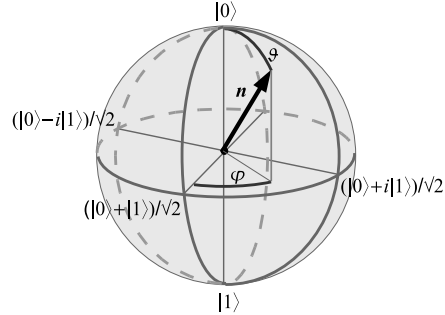,width=6cm}}
\caption{The Bloch sphere.} 
\label{fig:bloch}
\end{figure}

The discussion of such systems is made particularly transparent by
introducing the concept of the Bloch sphere\index{Bloch sphere} 
(actually, a ball). The
state (pure or mixed) of such a system is represented by a density
matrix: a $2\times 2$, hermitian, and positive definite operator with unit
trace. Any such operator can be written in the form
\begin{equation}\label{blochrep}
\rho=\frac{1}{2}(\mathbb{I}+\bm{n}\cdot\bm{\sigma}),
\end{equation}
where $\bm{n}$ is a real three-dimensional vector with $n=|\bm{n}|\le
1$, $\mathbb{I}$ is the 
unit operator, and $\bm{\sigma}$ is the vector of Pauli matrices in
the basis $|0\rangle,|1\rangle$. 
Thus, the state of a two-level system is represented in a unique way
by a unit ball in a three-dimensional real space (Fig.~\ref{fig:bloch}).
Since 
$\mathrm{Tr}\,\sigma_{i}=0$, $\mathrm{Tr}\,\mathbb{I}=2$, and
$(\bm{n}\cdot\bm{\sigma})^{2}=n^{2}\mathbb{I}$, one finds 
$\mathrm{Tr}\rho^{2}=n^{2}$, so that pure states (represented by
projectors) correspond to unit vectors $\bm{n}$ and are mapped to
the surface of the ball. It is easy to see that the point with
spherical coordinates $(\vartheta,\varphi)$ represents the state
vector
$\cos(\vartheta/2)|0\rangle+e^{i\varphi}\sin(\vartheta/2)|1\rangle$.

The electric field of the laser beam is
conveniently written as $\mathcal{E}(t)\cos(\omega t-\phi)$, where $\omega$
is the frequency of the light, $\phi$ is the phase of the pulse 
and $\mathcal{E}(t)$ is the envelope of
the pulse, varying slowly compared to the oscillations of the optical
field. The state of the system $|\Psi\rangle$
evolves according to the Schr{\"o}dinger equation
\begin{equation}\label{schroed}
i\hbar\frac{d}{dt}|\Psi\rangle=H|\Psi\rangle,
\end{equation}
with the Hamiltonian
\begin{equation}\label{ham-2level}
H=E|1\rangle\!\langle 1|
+\mathcal{E}(t-t_{\mathrm{a}})\cos[\omega(t-t_{\mathrm{a}})-\phi] 
\left( \mu |0\rangle\!\langle 1|+\mu^{*} |1\rangle\!\langle 0| \right),
\end{equation}
where $E$ is the energy of the interband transition in the QD, $\mu$
is the off-diagonal (interband) matrix element of the electric dipole
moment (between the two relevant states),  and we allow the pulse to
arrive at an arbitrary time $t_{\mathrm{a}}$. By a proper choice of
the phase of the state $|1\rangle$ it is possible to have $\mu$ real,
which will be assumed in the following. We will use the shorthand
notation $f(t)=\mu\mathcal{E}(t)$.

In the absence of the driving, the phase of the state $|1\rangle$
rotates with the frequency $E/\hbar$ which is of the order of
fs$^{-1}$. One gets rid of this trivial fast dynamics by describing
the system in the frame rotating with the same frequency. This is
similar to a transition to the interaction picture except that, for
practical reasons, it is convenient to perform the transformation with
the laser frequency $\omega$ instead of the system frequency $E/\hbar$
(these two frequencies are close to each other). Thus, we perform the
transformation 
\begin{displaymath}
|\tilde{\Psi}\rangle=U_{\mathrm{r}}|\Psi\rangle,
\quad U_{\mathrm{r}}=\exp(i\omega t|1\rangle\!\langle 1|).
\end{displaymath}
Using Eq.~(\ref{schroed}) one easily finds the evolution equation for
the redefined states in the form 
$i\hbar|\dot{\tilde\Psi}\rangle=\tilde{H}|\tilde\Psi\rangle$, with the
Hamiltonian
\begin{eqnarray}\label{ham-2level-rf}
\tilde{H}& = & -\hbar\omega |1\rangle\!\langle 1|
+U_{\mathrm{r}}HU_{\mathrm{r}}^{\dag} \\
\nonumber
& = & -\Delta|1\rangle\!\langle 1|
+\frac{1}{2}f(t-t_{\mathrm{a}})
\left[ 
\left( e^{-i\omega t_{0}-i\phi}+e^{-i\omega(2t-t_{0})+i\phi}\right)
|0\rangle\!\langle 1|
\phantom{\frac{1}{2}}\right.\\
\nonumber 
&&\left.
+\left( e^{i\omega t_{0}+i\phi}+e^{i\omega(2t-t_{0})-i\phi}\right)
|1\rangle\!\langle 0| \right],
\end{eqnarray}
where $\Delta=\hbar\omega-E$ is the detuning of the laser beam from
the transition energy.

The next essential step is to note that the natural frequency scale of
the system evolution is set by the detuning $\Delta$ and by the pulse amplitude
$f(t)$ and is many orders of magnitude smaller than the optical frequency
$\omega$. Therefore, the quickly oscillating terms 
($\sim e^{2i\omega t}$) are strongly off-resonant and will have very
little impact on the system evolution. Therefore, they can be
neglected, which leads to the rotating wave approximation (RWA)
\cite{scully97}.\index{rotating wave approximation}\index{RWA}
As a result, one obtains the following RWA Hamiltonian
\begin{eqnarray}\label{ham-2level-rwa}
H_{\mathrm{RWA}} & = & -\Delta|1\rangle\!\langle 1|
+\frac{1}{2}f(t-t_{\mathrm{a}})
\left( e^{-i(\omega t_{0}+\phi)}|0\rangle\!\langle 1|
+ e^{i(\omega t_{0}+\phi)} |1\rangle\!\langle 0| \right), \\
&=& -\Delta|1\rangle\!\langle 1|
+\frac{1}{2}f(t-t_{\mathrm{a}})\hat{\bm{u}}\cdot\bm{\sigma},
\nonumber
\end{eqnarray}
where $\hat{\bm{u}}=
[\cos(\omega t_{\mathrm{a}}+\phi),\sin(\omega t_{\mathrm{a}}+\phi),0]$.

\subsection{Pulse area dependent Rabi oscillations}
\label{sec:two-level-rabi}

Let us start with applying the formalism to the Rabi oscillations\index{Rabi oscillations} described in
Sec.~\ref{sec:exp}. First, assume that the pulse is resonant with the
optical transition in the QD, that is, $\Delta=0$. In this case, the
evolution operator generated by the RWA Hamiltonian
(\ref{ham-2level-rwa}) can be found analytically. Indeed,
in the resonant case the Hamiltonians at different times
commute with one another, so that the evolution operator may be
written as 
\begin{equation}
\label{U}
U(t)=\exp\left[ \frac{1}{2}\Phi(t)\hat{\bm{u}}\cdot\sigma \right] 
=\cos\frac{\Phi(t)}{2}\mathbb{I}-i\sin\frac{\Phi(t)}{2}
\hat{\bm{u}}\cdot\bm{\sigma},
\end{equation}
where 
$\Phi(t)=\int_{t_{0}}^{t}d\tau f(\tau)$,
$t_{0}$ is the
initial time of the evolution,
and the last identity is easily proven by expanding the exponent in a
series and collecting the odd- and even-order terms using the identities
$(\hat{\bm{u}}\cdot\bm{\sigma})^{2m}=\mathbb{I}$ 
and $(\hat{\bm{u}}\cdot\bm{\sigma})^{2m+1}=\hat{\bm{u}}\cdot\bm{\sigma}$
for any natural number $m$. We will always assume that this time is
before any pulses were switched on, so that one can set
$t_{0}\to -\infty$. The value of $\alpha\equiv \Phi(\infty)$ is called
the pulse area and determines the unitary transformation of the system
state performed by the complete pulse. One speaks of $\pi$-pulses,
$\pi/2$ pulses etc., referring to the value of the pulse area.

\begin{figure}[tb]
\begin{center}
\parbox{7cm}{\epsfig{figure=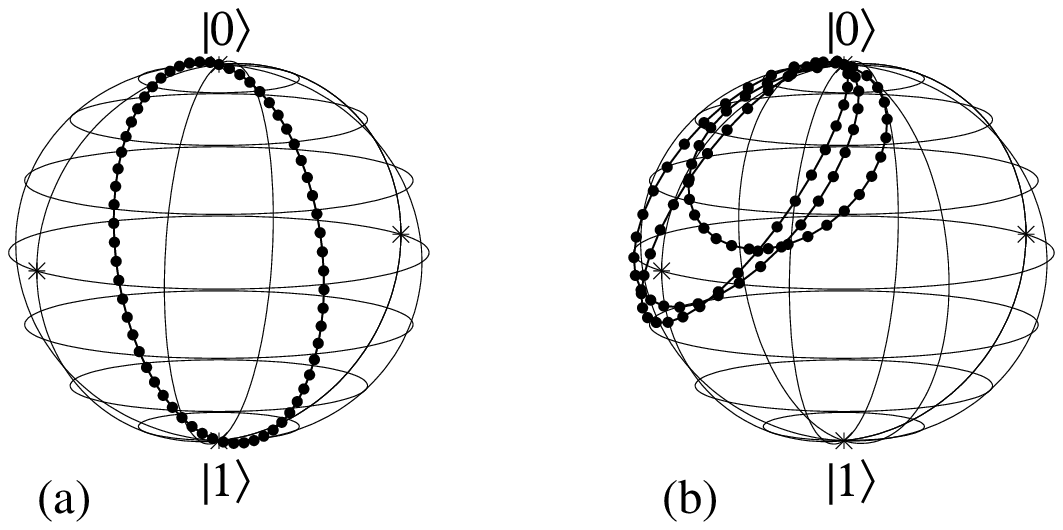,width=7cm}}
  \parbox{4cm}{\epsfig{figure=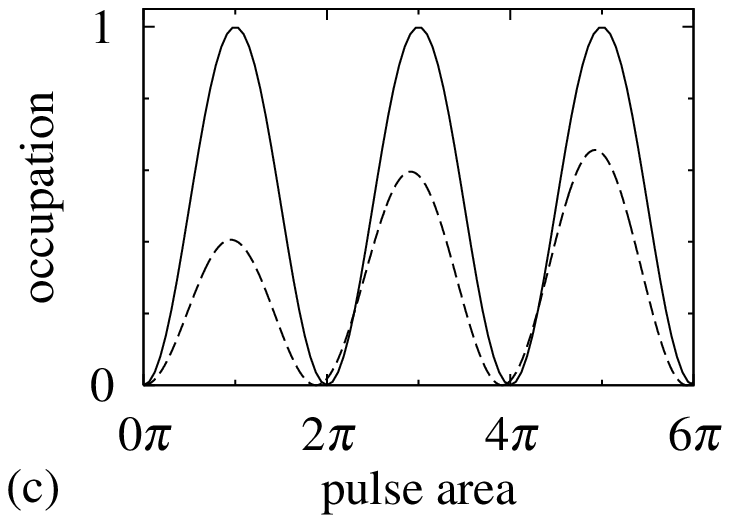,width=4cm}}
\end{center}
\caption{
(a) Pulse-area dependent Rabi oscillations at the resonance. The
points represent final system states for different pulse areas
$\alpha\in(0,2\pi)$ with a 
step of $\Delta\alpha=0.1$. (b) Pulse-area dependent Rabi oscillations
off resonance. The points represent final system states for 
different pulse areas $\alpha\in(0,6\pi)$ with a step of
$\Delta\alpha=0.2$. (c) The 
occupation of the state $|1\rangle$ as a function of the pulse area for
resonant (solid) and non-resonant (dashed) driving.}
\label{fig:rabi}
\end{figure}

Starting from the ground system state $|0\rangle$ one obtains, after
switching the pulse off, 
\begin{equation}
\label{after-pulse}
|\tilde\Psi\rangle=U(\infty)|0\rangle=
\cos\frac{\alpha}{2}|0\rangle-ie^{i\phi}\sin\frac{\alpha}{2}|1\rangle.
\end{equation}
The occupation of the state $|1\rangle$ is, therefore,
\begin{displaymath}
|\langle 1|\Psi\rangle|^{2}=|\langle 1|\tilde\Psi\rangle|^{2}
=\sin^{2}\frac{\alpha}{2}
\end{displaymath}
and indeed oscillates between 0 and 1 as a function of the pulse area
$\alpha$. The final system states for a set of values of $\alpha$ in
this resonant case are shown in Fig.~\ref{fig:rabi}a. Let us note here
that the final state is determined by a simple function of just one
quantity, the pulse area, and is independent of any details of the
pulse shape. This fact is known as the area theorem\index{area theorem}. 

Off resonance (for $\Delta\neq 0$), the evolution can be found in a
closed analytical 
form only for rectangular pulse envelopes $f(t)$ \cite{allen87}. Instead of
performing this simple exercise, let us look at the evolution for a
family of Gaussian pulses
\begin{equation}\label{gaussian-pulse}
f(t)=\frac{\hbar\alpha}{\sqrt{2\pi}\tau_{\mathrm{p}}}
e^{\frac{1}{2}\left(\frac{t}{\tau_{\mathrm{p}}}  \right)^{2} }
\end{equation}
obtained by (also very simple) numerical integration of the
Schr{\"o}dinger equation~(\ref{schroed}), shown in Fig.~\ref{fig:rabi}b
(see also Ref.~\refcite{vasconcellos06}). In the
detuned case, the system does not reach the $|1\rangle$
state. Instead, the state is rotated around a tilted axis and the
final state does not show periodicity as a function of the pulse
area. The difference between the resonant and non-resonant case is
also visible in the dependence of the final occupation of the state
$|1\rangle$ (Fig.~\ref{fig:rabi}c). 

\subsection{Time-domain interference}
\label{sec:2level-interf}

Another experiment that can easily be explained based on the two level
model is that of time domain interference. As discussed in
Sec.~\ref{sec:exp}, such experiments are performed with two
laser pulses selectively tuned to the exact resonance with one of the two 
fundamental optical transitions. The pulses are generated by splitting
a single laser pulse and delaying one part with respect to the
other. In this way, the two pulses are phase-locked, i.e., their
relative phase is definite and determined by the delay time $\tau$.
The first pulse arrives at $t_{\mathrm{a}}=0$ 
and prepares the initial superposition
state. In a usual two-slit (space-domain) experiment, this would correspond 
to splitting the particle path.
The second pulse arrives at $t_{\mathrm{a}}=\tau$. 
This pulse plays the role of ``beam merger''
providing, at the same time, a phase shift between the ``paths''.

The Hamiltonian describing the system driven by the two pulses
is obtained by an obvious generalization of
Eq.~(\ref{ham-2level-rwa}). 
\begin{equation}
\label{ham-interf}
H =   
\frac{1}{2}f_{1}(t)(|0\rangle\!\langle 1|+|1\rangle\!\langle 0|)
 +\frac{1}{2}f_{2}(t-\tau)
  (e^{-i\omega\tau}|0\rangle\!\langle 1| e^{i\omega\tau}|1\rangle\!\langle 0|),
\end{equation}
where $f_{i}(t)$ are the envelopes of the two
pulses and we have set $\phi=0$. The pulses do not overlap in time so
that the evolution can be split into two independent stages.

In an experiment, 
the system is initially in the state $|0\rangle$.
The first pulse is a $\pi/2$ pulse that performs the
transformation 
$U_{1}=(\mathbb{I}-i\sigma_{x})/\sqrt{2}$.
This pulse leaves the system in the equal superposition state 
\begin{equation}\label{psi}
|\psi\rangle=\frac{|0\rangle-i|1\rangle}{\sqrt{2}}. 
\end{equation}
The second pulse is again a $\pi/2$ pulse,
\begin{equation}\label{pulse-s2}
U_{2}=\frac{1}{\sqrt{2}}\left(  
\mathbb{I}-i\hat{\bm{n}}\cdot\bm{\sigma} 
\right),
\end{equation}
where $\hat{\bm{n}}=[\cos\omega\tau,\sin\omega\tau,0]$, as follows
from the general discussion in Sec.~\ref{sec:2level-general}.
After this pulse, the average number of excitons\index{exciton} 
in the dot is 
\begin{displaymath}
N(\phi)=|\langle 1|U_{\mathrm{2}}|\psi\rangle|^{2}
=\frac{1}{2}\left(  1-\cos\omega\tau \right),
\end{displaymath}
and changes periodically between $N_{\mathrm{min}}=0$ and
$N_{\mathrm{max}}=1$ as a function of the delay time $\tau$,
thus producing an interference pattern.

The quality of the interference pattern is quantified in terms of
visibility of interference fringes
\begin{displaymath}
{\cal V}=\frac{N_{\mathrm{max}}-N_{\mathrm{min}}}
{N_{\mathrm{max}}+N_{\mathrm{min}}}.
\end{displaymath}
The amplitude
(visibility) of the fringes in an ideal experiment is ${\cal V}=1$
for $\pi/2$ pulses, i.e., for an equal superposition state in between
the pulses. Otherwise, some \textit{a
priori} information on the superposition state can be inferred and the
visibility is reduced. Reduction of visibility occurs also if some
information on the system state between the pulses (analogous to
\textit{which way} information) has been extracted either
intentionally, in a controlled experiment (see
Sec.\ref{sec:biex-complem}), or as a result of uncontrolled dephasing
\cite{machnikowski06c}. 


\section{Beyond the two levels: optically driven evolution in a
biexciton system}\label{sec:biex}

The quantum control of a two-level system formed by the ground state
and a single exciton\index{exciton} exploits only small part of the possibilities
offered by a QD. Much more interesting physics, as well as potential
applications, becomes available if all the states   are involved in the
dynamics. Below we study two quantum optical schemes involving the
confined biexciton\index{biexciton} system. First, the theory of the experimentally
observed two-photon Rabi oscillations is given \cite{stufler06}. Then,
a theoretical 
proposal for an experiment highlighting the role of
quantum complementarity in time-domain interference is described
\cite{machnikowski06c}. 

\subsection{Two-photon Rabi oscillations}
\index{Rabi oscillations!two-photon}
If we apply circularly polarized excitation to a QD in the ground
state we can only create single excitons\index{exciton}, as follows from the
selection rules discussed in Sec.~\ref{sec:qd} and from the Pauli
exclusion principle. 
By using linear polarization, which is a superposition
of two equal $\sigma_{+}$ and $\sigma_{-}$ polarized components, 
one enables also biexciton\index{biexciton} generation. 
In this case, the
ground state is coupled to the biexciton\index{biexciton} state via both exciton\index{exciton}
states. Here we will consider a situation in which the laser frequency
is tuned to the two-photon resonance with the biexciton\index{biexciton}, while the
single-exciton\index{exciton} states are detuned by $\Delta=E_{\mathrm{B}}/2$, as shown in
Fig.~\ref{fig:biex-diagram}b.

In the rotating wave approximation, the Hamiltonian of the biexcitonic
system described above can be written in the form
\begin{equation}\label{ham-twophot}
    H = \frac{E_{B}}{2}
        \left( |+\rangle\!\langle +|+|-\rangle\!\langle -| \right)
 +\frac{f(t)}{2\sqrt{2}}\left[  
(|\mathrm{g}\rangle+|\mathrm{XX}\rangle)(\langle +|+\langle -|) +\mathrm{H.c.} \right],
\end{equation}
where 
$|+\rangle$,$|-\rangle$ and $|\mathrm{XX}\rangle$ are the $\sigma_{+}$,
$\sigma_{-}$ and biexciton\index{biexciton} states, respectively.
Thus, the laser pulse couples only one (bright) combination of the
states ground and biexciton\index{biexciton} states,
$|\mathrm{B}\rangle=(|\mathrm{g}\rangle+|\mathrm{XX}\rangle)/\sqrt{2}$ to one ($\pi_{x}$--polarized)
single exciton\index{exciton} state $|\mathrm{X}\rangle=(|+\rangle+|-\rangle)/\sqrt{2}$. 

The system evolution is easily described in the new basis
$|Y\rangle=(|+\rangle-|-\rangle)/\sqrt{2}$,$|\mathrm{D}\rangle=(|\mathrm{g}\rangle-|\mathrm{XX}\rangle)/\sqrt{2}$,
$|\mathrm{X}\rangle$,$|\mathrm{B}\rangle$,
where the first two states are the (decoupled) $\pi_{y}$-polarized 
exciton\index{exciton} state and the dark combination of the
states ground and biexciton\index{biexciton} states. In this basis, 
the Hamiltonian (\ref{ham-twophot}) can be written as
\begin{equation}\label{ham-red}
    H=\left(\begin{array}{cccc}
      E_{B}/2 & 0 & 0 & 0 \\
      0 & 0 & 0 & 0 \\
      0 & 0 & E_{B}/2 & f(t)/\sqrt{2} \\
      0 & 0 & f(t)/\sqrt{2} & 0 \\
    \end{array}  \right),
\end{equation}
which corresponds to the detuned rotation in the invariant subspace spanned by
the $x$--polarized exciton\index{exciton} state $|X\rangle$ and the state
$|\mathrm{B}\rangle$, while the first two states are decoupled
and undergo only a trivial evolution. 
The initial (ground) state is now written as
$|\mathrm{g}\rangle=\frac{|\mathrm{D}\rangle+|\mathrm{B}\rangle}{\sqrt{2}}.$

Let us consider the instantaneous (for fixed $t$) eigenstates of the
Hamiltonian (\ref{ham-red}) as a function of $f(t)$. 
For a large binding energy $E_{\mathrm{B}}$ 
the two branches belonging to the non-trivial two-dimensional 
invariant subspace are always widely separated. Thus, under the action
of a sufficiently slowly varying laser pulse the system undergoes an
adiabatic evolution with the pulse envelope $f(t)$ playing the role of a
slowly varying parameter of the Hamiltonian.
Therefore, one can assume \cite{messiah66} that at each time $t$ the
state corresponds to the adiabatic (instantaneous) eigenstate of the
Hamiltonian (\ref{ham-red}), i.e.,
\begin{displaymath}
    |\psi(t)\rangle
     =\frac{|\mathrm{D}\rangle +\left[
            c_{-}(t)|\mathrm{X}\rangle+c_{+}(t)|\mathrm{B}\rangle
     \right] e^{-i\Lambda(t)}}{\sqrt{2}},
\end{displaymath}
where
\begin{displaymath}
    c_{\pm}(t)  =  \frac{1}{\sqrt{2}}\left(
        1\pm\frac{E_{B}}{\sqrt{(E_{B})^{2}+8f^{2}(t)}}
		 \right)^{1/2},
\end{displaymath}
and
\begin{equation}\label{lambda}
    \Lambda(t)=\frac{1}{4\hbar}\int_{-\infty}^{t}d\tau
        \left( E_{B}-\sqrt{(E_{B})^{2}+8f^{2}(\tau)}\right).
\end{equation}

After the laser pulse has been switched off, the state becomes
\begin{equation}\label{final}
|\psi\rangle = 
\frac{|\mathrm{D}\rangle+e^{-i\Lambda(\infty)}|\mathrm{B}\rangle}{\sqrt{2}}
= e^{-i\Lambda(\infty)/2}\left( 
 \cos\frac{\Lambda(\infty)}{2}|\mathrm{g}\rangle
	+ \sin\frac{\Lambda(\infty)}{2}|\mathrm{XX}\rangle\right).
\end{equation}
and coincides with the biexciton\index{biexciton} state for $\Lambda(\infty)=\pi$.
Thus, the system effectively performs Rabi oscillations
\index{Rabi oscillations!two-photon} between the
ground state and the biexciton\index{biexciton} state, with the 
occupation of the biexciton\index{biexciton} state given by
\begin{equation}\label{N2}
    N_{\mathrm{XX}}=
|\langle \mathrm{XX}| \psi\rangle|^{2}=\sin^{2}\frac{\Lambda(\infty)}{2}.
\end{equation}

For $\theta\ll \tau_{0}\Delta E_{B}/\hbar$
one has
\begin{displaymath}
\Lambda(\infty)\approx \frac{1}{\hbar\Delta E_{B}}
 \int_{-\infty}^{\infty}dt f^{2}(t)
=\frac{4\hbar\mathrm{Arcosh}\sqrt{2}}{\pi^{2}\Delta E_{B}\tau_{0}}\theta^{2}
\end{displaymath}
and the biexciton\index{biexciton} occupation $N_{\mathrm{XX}}$ grows as $\theta^{4}$, i.e.,
proportional to the square of the pulse intensity, as expected for
a two-photon process (see Fig.~\ref{fig:Rabi}).

It should be noted that the pulse area appearing as the parameter
of the standard Rabi oscillations is now replaced by the adiabatic
dynamical phase $\Lambda$, which is a nontrivial functional of the
pulse shape [Eq.~(\ref{lambda})]. 
In this way, the simple universal dynamics off a resonantly
driven two level system, described by the area theorem, is
replaced by a more complicated pattern of oscillations, depending
on the detuning parameter $\tau_{\mathrm{p}}E_{\mathrm{B}}$ but also
on the exact pulse shape (e.g. Gaussian vs. sech \cite{stufler06}).
The absence of any single exciton\index{exciton} occupation in the final state is
obviously due to the fact that 
the adiabatic limit corresponds to weak and long (i.e., spectrally narrow)
pulses and the excitation of
single-exciton states becomes forbidden by energy
conservation. Nonetheless, during the evolution the
single-exciton states are also occupied. 

\subsection{Quantum complementarity in time-domain interference
experiments} 
\label{sec:biex-complem}

This Section discusses the essential modification to time domain
interference experiments (Sec.~\ref{sec:2level-interf}) 
that allows one to attain
partial information on the state of the system and to observe the
related visibility reduction of the interference pattern. 
First, however, a measure of the partial information is introduced and
the complementarity principle is formulated in a quantitative 
form \cite{jaeger95,englert96}. 

The notion of ``partial information'' is understood as follows. The
system (S) of interest is coupled to another \textit{quantum
probe} (QP) system and conditional dynamics of the latter is
induced, leading to correlations between the states of the systems S
and QP. Next, a measurement on QP is performed and its result is used
to infer the state of S, i.e., to predict the outcome of a subsequent
measurement on S. The probability of a correct prediction ranges from
1/2 (guessing at random in absence of any correlations) to 1 (knowing
for sure, when the systems are maximally entangled). 

Quantitatively, an intrinsic measure of information on
the system S extracted by QP is provided by the
\textit{distinguishability of states} \cite{jaeger95,englert96},
\begin{equation}\label{disting-def}
{\cal D}=2\left(p-\frac{1}{2}\right), 
\end{equation}
where $p$ is the probability a correct prediction for the state of S
maximized over all possible measurements on QP.
In this way, guessing at random and knowing for sure correspond to 
${\cal D}=0$ and ${\cal D}=1$, respectively. 
According to a general theory \cite{jaeger95,englert96},
the complementarity relation between the 
knowledge of the system state and the visibility of the fringes may be
written, using the distinguishability ${\cal D}$ as a measure of
information, in the quantitative form, 
\begin{equation}\label{ineq}
{\cal D}^{2}+{\cal V}^{2}\le 1.
\end{equation}
The equality holds for systems in pure states. 

In a QD, this formal scheme translates
naturally into the conditional dynamics of a biexcitonic system
(Fig.\ref{fig:biex-diagram}a), as described in Sec.~\ref{sec:exp}.
Thus, the exciton\index{exciton} addressed in the interference experiment
described in Sec.~\ref{sec:2level-interf} is, say, the $\sigma_{+}$ exciton\index{exciton}
(system S). The other degree of freedom of the biexciton\index{biexciton} system (the
$\sigma_{-}$ exciton\index{exciton}) will be used as the quantum probe.

\begin{figure}[tb]
\begin{center}
\parbox{3cm}{\epsfig{figure=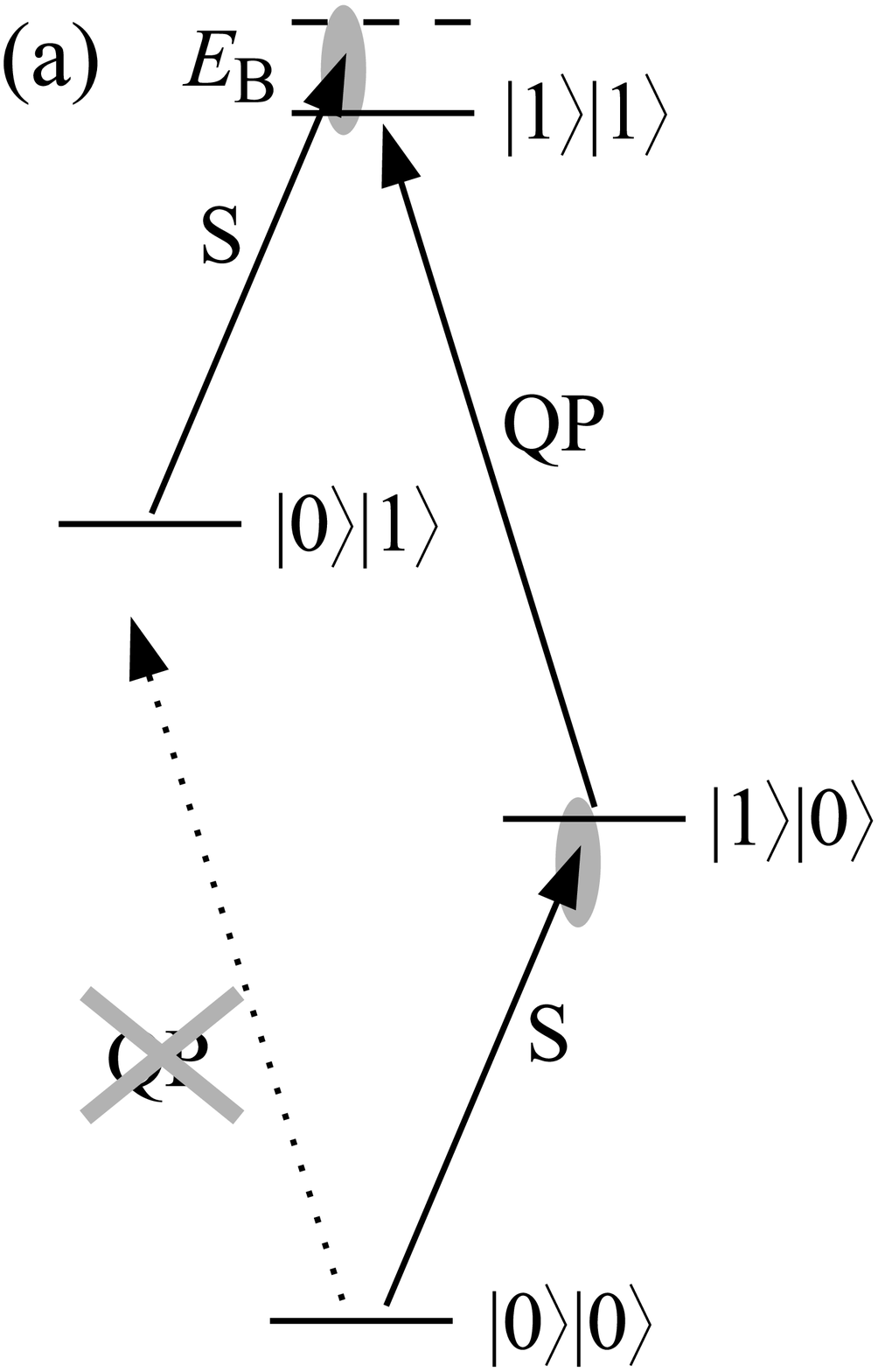,width=3cm}}
  \hspace*{1.5cm}
  \parbox{5.1cm}{\epsfig{figure=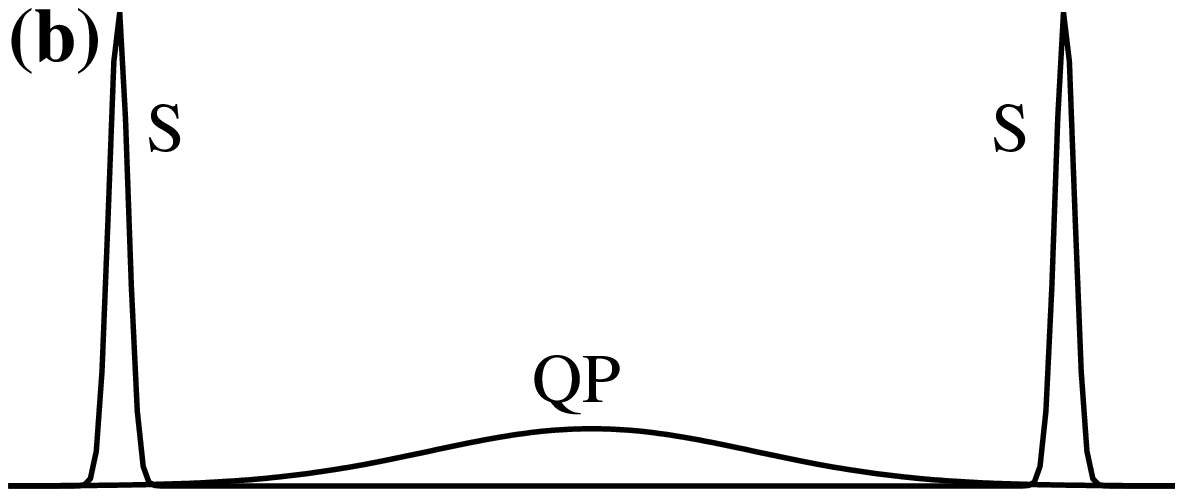,width=5.1cm}}
\end{center}
\caption{(a) The diagram of energy levels and
transitions in the system, assuming that the excitons\index{exciton}
are confined in 
neighboring dots with different transition energies. The
pulses inducing transitions on the system S have broad spectrum so
that both transitions are possible. The pulse acting on
the QP system is spectrally selective and tuned to the biexcitonic
transition, so that the single-exciton transition is
energetically forbidden in this subsystem. 
(b) The sequence of pulses used in the experiment.}
\label{fig:complem}
\end{figure}

We will use a tensor product notation with $|0\rangle$ and $|1\rangle$
denoting the absence and presence of the respective exciton\index{exciton}, as
previously, with the
interfering system (S) always to the left. In the rotating 
basis with respect to both subsystems,
the RWA Hamiltonian for the biexciton\index{biexciton} system is 
\begin{equation}\label{ham0}
H =  H_{1}\otimes\mathbb{I}
+ \Delta |1\rangle\!\langle 1|\otimes 
|1\rangle\!\langle 1|
+\frac{1}{2}f_{\mathrm{QP}}(t-\tau_{\mathrm{QP}})
  \mathbb{I}\otimes(|0\rangle\!\langle 1| +\mathrm{H.c.})
\end{equation}
where $\Delta$ is the bi-exciton energy shift and $E_{\mathrm{QP}}(t)$
is the envelope of the pulse coupled to the QP exciton\index{exciton}. 
Here the first 
term denotes the Hamiltonian (\ref{ham-interf}) and
corresponds to the pulse sequence of the interference
experiment described in Sec.~\ref{sec:2level-interf}, 
the second one accounts for the bi-excitonic energy shift
and the third term describes the action of the pulse coupled to the
second (QP) exciton\index{exciton} and spectrally tuned to the exciton\index{exciton}-biexciton
transition. This pulse arrives at $t=\tau_{\mathrm{QP}}$, between the
other two pulses (that is, $0<t_{\mathrm{QP}}<\tau$), and
will induce the conditional dynamics. 
Its phase is irrelevant and will be assumed 0. The
structure of system excitations and the sequence of pulses are
shown in Fig.~\ref{fig:complem}.

Assume that the exciton\index{exciton} (system S) 
is in the equal superposition state (\ref{psi}).
The probability of correctly
guessing the result of a measurement in the $|0\rangle,|1\rangle$ basis
without any additional information is obviously 1/2. 
Now, we can correlate this excitonic system with the other one (QP;
initially in the state $|0\rangle$). 
To this end, one applies a selective (spectrally narrow) pulse with
the area $\alpha$ (the arrow labelled `QP' in Fig.~\ref{fig:complem}a).
This rotates the state of the second subsystem according to
Eq.~(\ref{U}) (with $\hat{\bm{n}}=[1,0,0]$) if and only if the first
system is in the state $|1\rangle$. The corresponding unitary
transformation of the compound system is 
\begin{displaymath}
U_{\mathrm{QP}}=|0\rangle\!\langle 0|\otimes \mathbb{I}
+|1\rangle\!\langle 1|\otimes 
\left( \cos\frac{\alpha}{2}\mathbb{I}
-i \sin\frac{\alpha}{2}\sigma_{x} \right),
\end{displaymath}
and takes the state $|\psi\rangle$ into 
\begin{displaymath}
|\psi'\rangle=\frac{1}{\sqrt{2}}|0\rangle\otimes|0\rangle
-\frac{i}{\sqrt{2}}|1\rangle 
\otimes\left(\cos\frac{\alpha}{2}|0\rangle
-i\sin\frac{\alpha}{2}|1\rangle\right). 
\end{displaymath}
For $\alpha=\pi$, this pulse performs
a CNOT-like transformation \cite{nielsen00} on the biexcitonic
system. As a result, the 
total system is in the maximally entangled state 
$(|0\rangle|0\rangle-i|1\rangle|1\rangle)/2$ and
a measurement on the QP system uniquely determines the state of
the system S. Hence, due to quantum correlations between the systems,
complete information on the state of S has been extracted to QP.
On the other hand, if the
biexciton is excited with a pulse with area $\alpha<\pi$ 
the correlation between the subsystems is weaker and a measurement on
QP cannot fully determine the state of S, although the
attained information may increase the probability for correctly predicting the
result of a subsequent measurement on S. According to the discussion
above, this means that partial information on the state of S is available.

In order to find the distinguishability measure in the biexciton\index{biexciton}
scheme discussed here, we write the density matrix of the total system
corresponding to the state $|\psi'\rangle$,
\begin{equation}\label{dm-full}
\varrho=\frac{1}{2}\sum_{nm}|n\rangle\!\langle m|\otimes\rho_{nm},
\end{equation}
with 
$\rho_{00} = |0\rangle\!\langle 0|$, 
$\rho_{11} =  \frac{1}{2}\left( \mathbb{I}+\cos\alpha\ \sigma_{z}
-\sin\alpha\ \sigma_{y} \right)$,  
$\rho_{01}=\rho_{10}^{\dag} 
= i\cos\frac{\alpha}{2}|0\rangle\!\langle 0|
-\sin\frac{\alpha}{2}|0\rangle\!\langle 1|$.
Note that $\rho_{00}$ and
$\rho_{11}$ (but not $\rho_{01}$) are density matrices.

According to the general theory \cite{jaeger95,englert96}, 
the best chance for correctly guessing the
state of S results from the measurement of the
observable $\rho_{00}-\rho_{11}$ and the probability of the correct
prediction is then 
$p=\frac{1}{2}+\frac{1}{4}\mathrm{Tr} |\rho_{00}-\rho_{11}|$
where
$|\rho|$ is the modulus of the operator $\rho$. Using the
explicit forms of the density matrices $\rho_{00},\rho_{11}$ and the
definition (\ref{disting-def}) one finds
for the distinguishability in our case
\begin{equation}\label{disting}
{\cal D}=\frac{1}{2}\mathrm{Tr} \left|\rho_{00}-\rho_{11}\right|
=\left|\sin\frac{\alpha}{2}\right|.
\end{equation}
Thus, the amount of information on the system S accessible via
a measurement on QP increases from 0 (no QP pulse at all) to 1 (for a
$\pi$ pulse).

Next, we study the effect of extracting the which path information 
on the interference fringes. In the state (\ref{dm-full}),
the reduced density matrix of the subsystem S is 
$\rho_{\mathrm{S}}=\mathrm{Tr}_{\mathrm{QP}} \varrho=
(1/2)\left( \mathbb{I}-\cos(\alpha/2)\sigma_{y} \right)$.
Upon applying the second pulse of the interference experiment scheme,
namely the unconditional $\pi/2$ pulse [Eq.~(\ref{pulse-s2})],
the average number of excitons in the dot is 
\begin{displaymath}
N(\phi)=
\langle 1|U_{\mathrm{2}}\rho_{\mathrm{S}}U_{\mathrm{2}}^{\dag}|1\rangle
=\frac{1}{2}\left(  1-\cos\frac{\alpha}{2}\cos\phi \right).
\end{displaymath}
Now, the average occupation oscillates between the limiting values 
$(1\pm|\cos\alpha|)/2$ and the visibility of the fringes is
\begin{equation}\label{visib}
{\cal V}=|\cos(\alpha/2)|. 
\end{equation}
Comparing Eq.~(\ref{visib}) with Eq.~(\ref{disting}) it is clear that the
more certain one is whether the exciton\index{exciton} \textit{is there} or
\textit{not} ($\cal D$ increases), the less clear the interference fringes
become ($\cal V$ decreases). 
Quantitatively, the relation ${\cal D}^{2}+{\cal V}^{2}=1$
holds which is consistent with the complementarity relation (\ref{ineq}).
It can be shown that, in the presence of coupling to the environment, 
where both subsystems are in mixed states due to dephasing, this
relation will turn into inequality \cite{machnikowski06c}.
Let us notice that the impact of the which path information on
interference fringes is the same no matter
whether the QP subsystem is measured before or after generating
and detecting the interference fringes and even whether it is measured
at all. 

The time-domain manifestation of quantum complementarity discussed here 
not only broadens
the class of experiments in which fundamental aspects of the quantum
world may be tested but also has the advantage of being independent
of the position-momentum (Heisenberg's) uncertainty that has been 
historically tied
to the space-domain discussions of complementarity \cite{wooters79}. 
In fact, it is independent of any uncertainty principles whatsoever.
Indeed, the only two quantities which are measured 
in the time-domain interference experiment are the occupations
of two different excitons. These quantities refer to \textit{different
subsystems} and, therefore, are obviously \textit{commuting} and
simultaneously measurable. Although the quantum probe exciton\index{exciton} is
created in a way that correlates it with the presence of the other exciton\index{exciton}
(S), this cannot be interpreted as a (projective) measurement on S, 
since the QP exciton\index{exciton} is definitely a quantum (microscopic) system and
not a classical, macroscopic measurement device. Thus, the
experimental procedure described in this section demonstrates quantum
complementarity in its pure form, involving only the notion of
\textit{information} on the system state and independent of any
uncertainty relations between non-commuting observables.

Further discussion of the concept of complementarity in the context of
optical experiments on QDs, including the analysis of its
feasibility in terms of the currently available experimental
techniques and of the parameters of the existing structures, as well
as the analysis of the role of dephasing is given in
Ref.~\refcite{machnikowski06c}. 

\section{Carrier-phonon interaction in quantum dots}\label{sec:phonons}

\index{phonon}This section presents the derivation of
the coupling constants between bulk phonon modes and
confined carriers in a semiconductor
\cite{mahan00,haken76}. We will restrict ourselves to the
deformation potential coupling, which is relevant for the discussion
in the following sections. The treatment of other couplings
(piezoelectric and polar) is reviewed in Ref.~\refcite{grodecka05a}.

Any crystal deformation leads to shifts of the conduction (c) and
valence (v) bands which are, to the leading order, proportional to
the relative volume change. The corresponding contribution to the
energy of electrons (e) and holes (h) in the long-wavelength limit
is
\begin{displaymath}
    H^{(\mathrm{DP})}_{\mathrm{e/h}} \equiv \pm\Delta E_{\mathrm{c/v}}
    =\mp\sigma_{\mathrm{e/h}}\frac{\delta V}{V},
\end{displaymath}
where $\sigma_{\mathrm{e/h}}$ are the deformation potential
constants for electrons and holes and $V$ is the unit cell volume.
Using the strain tensor $\hat{\sigma}$,
\begin{displaymath}
    \sigma _{ij}=\frac{1}{2}\left(
\frac{\partial u_i}{\partial r_j} +\frac{\partial u_j}{\partial
r_i}\right),
\end{displaymath}
one may write $H^{(\mathrm{DP})}_{\mathrm{e/h}}
        =\mp\sigma_{\mathrm{e/h}} \mathrm{Tr}\hat{\sigma}
        =\mp\sigma_{\mathrm{e/h}} \nabla\cdot\bm{u}(\bm{r})$,
where $\bm{u}(\bm{r})$ is the local displacement field. The
displacement is quantized in terms of phonons,
\begin{equation}\label{phonon}
    \bm{u}(\bm{r})=i\frac{1}{\sqrt{N}}\sum_{\bm{k}}
        \sqrt{\frac{\hbar}{2\rho V \omega_{\bm{k}}}}
        \hat{\bm{e}}_{\bm{k}}\left(
            b_{\bm{k}}+b^{\dag}_{-\bm{k}}
        \right)e^{i\bm{k}\cdot\bm{r}},
\end{equation}
where $\omega_{\bm{k}}$ is the frequency for the wave vector
$\bm{k}$, $\hat{\bm{e}}_{\bm{k}}=-\hat{\bm{e}}_{-\bm{k}}$ is
the corresponding real unit polarization vector, and $\rho$ is the
crystal density. Only the longitudinal branch contributes to
$\nabla\cdot\bm{u}$ in (\ref{phonon}) and the final interaction
Hamiltonian in the coordinate representation for carriers is
\begin{equation}\label{int-DP-r}
    H^{(\mathrm{DP})}_{\mathrm{e/h}}
        =\pm\sigma_{\mathrm{e/h}}\frac{1}{\sqrt{N}}\sum_{\bm{k}}
        \sqrt{\frac{\hbar k}{2\rho V \omega_{\bm{k}}}}
        \left( b_{\bm{k}}+b^{\dag}_{-\bm{k}}
        \right)e^{i\bm{k}\cdot\bm{r}}.
\end{equation}

In the second quantization representation with respect to the
carrier states this reads
\begin{eqnarray}
    H_{\mathrm{e/h}}^{(\mathrm{DP})}
  & = & \sum_{nn'}\langle n|V(\bm{r})|n'\rangle a_{n}^{\dag}a_{n'}
    = \sum_{nn'}
    \int_{-\infty}^{\infty} d^3r\psi _n^* (\bm{r}) V(\bm{r})
    \psi_{n'}(\bm{r}) a_{n}^{\dag}a_{n'} \nonumber \\
  & = & \frac{1}{\sqrt{N}}\sum_{\bm{k}nn'}a_{n}^{\dag}a_{n'}
  f_{\mathrm{e/h},nn'}^{(\mathrm{DP})}(\bm{k})
\left(b_{\bm{k}}+b_{-\bm{k}}^{\dag}\right),
\label{int-DP}
\end{eqnarray}
where
$f_{\mathrm{e/h},nn'}^{(\mathrm{DP})} (\bm{k})
    =\sigma_{\mathrm{e/h}}\sqrt{\frac{\hbar k}{2\rho V c}}
    \mathcal{F}_{nn'} (\bm{k})$,
with the formfactor
\begin{equation}\label{formfactor}
    \mathcal{F}_{nn'}(\bm{k})
       =\int_{-\infty}^{\infty} d^3\bm{r}\psi _n^*
        (\bm{r})e^{i\bm{k}\cdot\mathrm{r}}\psi_{n'}(\bm{r})
       =\mathcal{F}_{n'n}^{*}(-\bm{k}).
\end{equation}

While the common coefficient of the coupling Hamiltonian contains
the fundamental and material-dependent constants and reflects
general electrical and mechanical properties of the semiconductor
system, the formfactor (\ref{formfactor}) contains information
about the geometry of the confinement and the resulting properties
of wave functions. In this sense, it is the ``engineerable'' part
of the carrier-phonon coupling.

From the orthogonality of single-particle states one immediately has
$\mathcal{F}_{nn'}(0)=\delta_{nn'}$. If the wave functions are
localized at a length $l$, then the extent of the formfactor is
$\sim 1/l$. Thus, for carrier states localized in a QD over many
lattice sites and smooth within this range, the functions
$\mathcal{F}_{nn'}(\bm{k})$ will be localized in
$\bm{k}$--space very close to the center of the Brillouin zone.

As an example, let us consider a Gaussian wave function,
\begin{equation}
    \psi(\bm{r})=\frac{1}{\pi^{3/4}l_{z}l_{\bot}}
      \exp\left[-\frac{1}{2}\left(\frac{r_{\bot}}{l_{\bot}}\right)^{2}
        -\frac{1}{2}\left( \frac{z}{l_{z}}\right)^{2}\right],
\label{ground-state}
\end{equation}
where $r_{\bot}$ is the position component in the $xy$ plane and
$l_{\bot},l_{z}$ are the localization widths in-plane and in the
growth ($z$) direction. The corresponding formfactor is then
easily found to be
\begin{equation}\label{formf-expli}
    \mathcal{F}(\bm{k})=\exp\left[
        -\left( \frac{k_{\bot}l_{\bot}}{2}\right)^{2}
        -\frac{1}{2}\left( \frac{k_{z}l_{z}}{2}\right)^{2}\right].
\end{equation}

The formulas derived above describe the interaction in the
single-particle basis. However, most of the following deals with
excitonic states, i.e. states of confined electron-hole pairs
interacting by Coulomb potentials.
Both carriers forming the exciton\index{exciton} couple to phonons according to
Eq.~(\ref{int-DP}). For the further discussion in this chapter it is
sufficient to discuss the lowest
exciton state. It is reasonable to assume its wave function 
approximately in a product form \cite{jacak03b}, i.e., 
\begin{displaymath}
    |1\rangle
        =a_{\mathrm{e},0}^{\dag}a_{\mathrm{h},0}^{\dag}|0\rangle,
\end{displaymath}
where $a_{\mathrm{e/h},0}^{\dag}$ create an electron and a hole in
the ground confined state in the QD.
Then, one gets from Eq.~(\ref{int-DP}) the following coupling constant
between the confined exciton\index{exciton} and phonons
for the deformation potential interaction
\begin{equation}\label{cpl-X-1}
  g_{\bm{k}} =
    \sqrt{\frac{\hbar k}{2\rho V c_{\mathrm{l}}}}\left(
    \sigma_{\mathrm{e}}\mathcal{F}^{(\mathrm{e})}(\bm{k})
    -\sigma_{\mathrm{h}}\mathcal{F}^{(\mathrm{h})}(\bm{k})
    \right).
\end{equation}
Note that, due to different deformation
potential constants $\sigma_{\mathrm{e/h}}$, this
coupling does not vanish even if the electron and hole wave functions are
the same, leading to identical single-particle formfactors.

Thus, the exciton\index{exciton} (restricted to its ground state) 
driven by a laser field and interacting
with LA phonons via the deformation potential coupling is described by
the Hamiltonian
\begin{equation}\label{ham-X-ph}
H=H_{\mathrm{RWA}}
+\sum_{\bm{k}}\hbar\omega_{\bm{k}}b_{\bm{k}}^{\dag}b_{\bm{k}}
+|1\rangle\!\langle 1| \sum_{\bm{k}}g_{\bm{k}} (b_{\bm{k}}+b_{-\bm{k}}^{\dag}),
\end{equation}
where the first term describes the carrier subsystem and is given by
Eq.~(\ref{ham-2level-rwa}). 

The Hamiltonian (\ref{ham-X-ph}) is the basis for
the microscopic modeling of dephasing effects in QDs. It turns out that this
relatively simple model is very successful in reproducing the 
experimental data \cite{vagov03,vagov04}. Hence, it may serve
as a reliable starting point for describing the evolution of the
combined system of confined carriers and lattice modes. 

\section{Theoretical methods for carrier--phonon
kinetics}\label{sec:kinet}

In this section we will study a few theoretical methods that have
proven to be useful for the description of the carrier-phonon quantum
kinetics in QDs. This discussion will be limited to the simplest case
of a two-level system described by the Hamiltonian (\ref{ham-X-ph}).

\subsection{Exact solution for ultrafast excitations}
\label{sec:ultrafast}

Let us start with the case of an ultrafast excitation. A very short laser pulse 
prepares the system in a certain superposition (dependent on the pulse
phase and intensity) of the ground state $|0\rangle$
(no exciton\index{exciton}) and the single-exciton state
$|1\rangle$. By \textit{very short} we mean a pulse 
much shorter than the time scales of phonon
dynamics, so that the preparation of the initial state may be
considered instantaneous. This corresponds to the actual experimental
situation with pulse durations of the order of $100$ fs \cite{borri01}. On the
other hand, the pulse is long enough to assure a relatively narrow
spectrum and to prevent the population of higher confined levels
and excitation of optical phonons \cite{machnikowski04a}. In this
ultrafast limit the only role of the laser pulse is to prepare the
initial state, while the subsequent evolution is generated by a
time-independent Hamiltonian, 
\begin{equation}
\label{exact-ham}
H=-\Delta |1\rangle\!\langle 1|
+\sum_{\bm{k}}\hbar\omega_{\bm{k}}b_{\bm{k}}^{\dag}b_{\bm{k}}
+|1\rangle\!\langle 1| \sum_{\bm{k}}g_{\bm{k}} (b_{\bm{k}}+b_{-\bm{k}}^{\dag}),
\end{equation}
obtained by setting $H_{\mathrm{RWA}}=-\Delta|1\rangle\!\langle 1|$ in
Eq.~(\ref{ham-X-ph}).

In a superposition state created by the laser pulse
[Eq.~(\ref{after-pulse})] the inter-band component of the
electric dipole moment has a non-vanishing average value
oscillating at an optical frequency (hence referred to as
\textit{optical polarization}) \cite{schafer02}
which leads to the emission
of coherent electromagnetic radiation with
an amplitude proportional to the oscillating dipole moment. In an
unperturbed system (e.g., in an atom), the radiation
would be emitted over time of
the order of the lifetime of the superposition state, i.e., until the
system relaxes to the ground state due to radiative energy loss. 

In a semiconductor structure an additional effect, related to
carrier-phonon coupling, appears on a time scale much shorter than the
lifetime of the state.  
The last two terms in Eq.~(\ref{exact-ham}) describe a set of
harmonic oscillators which, in the presence of the exciton\index{exciton}, are
displaced by an external force proportional to the coupling constant
$g_{\bm{k}}$. 
This means that, due to the interactions between confined carriers and
lattice ions,  
the ground state of the lattice in the
presence of a charge distribution is different than in its
absence. As a result, after the creation of a confined exciton\index{exciton} 
the lattice 
relaxes to a new equilibrium, which is accompanied by the emission of
phonon wave packets \cite{jacak03b,vagov02a} that form a trace in the
macroscopic crystal distinguishing the exciton\index{exciton} state from an empty
dot. This information broadcast via emitted
phonons leads to a decay of the coherence of the superposition state
\cite{roszak06b} although the average
occupations of the system states remain unaffected (hence the process
is referred to as \textit{pure dephasing}).
Since coherent dipole radiation requires well-defined phase relations 
between the components
of a quantum superposition, the amplitude of this radiation, measured in
the experiment, gives access to the coherence properties of the
quantum state of confined carriers itself. The dephasing of the quantum
superposition is therefore directly translated into the decay of coherent
optical radiation from the system.

As discussed in Sec.~\ref{sec:phonons}, the carrier-phonon interaction term in
Eq.~(\ref{exact-ham}) is linear in phonon operators
and describes a shift of the lattice equilibrium induced by the
presence of a charge distribution in the dot. The stationary state of
the system corresponds to the exciton\index{exciton} and the surrounding coherent 
cloud of phonons representing the lattice distortion to the new
equilibrium. The transformation that creates the coherent cloud is
the shift 
$wb_{\bm{k}}w^{\dagger}=b_{\bm{k}}-f_{\bm{k}}/(\hbar\omega_{\bm{k}})$,
generated by the Weyl operator \cite{haken76,roszak06b}
\begin{equation}
\label{w}
w= 
\exp\left[\sum_{\bm{k}} 
\left( \frac{g_{\bm{k}}^{*}}{\hbar\omega_{\bm{k}}}b_{\bm{k}}^{\dag}
-\frac{g_{\bm{k}}}{\hbar\omega_{\bm{k}}} b_{\bm{k}} \right)\right].
\end{equation}
A straightforward calculation shows that the Hamiltonian (\ref{exact-ham})
is diagonalized by the unitary transformation 
$W=|0\rangle\!\langle 0|\otimes \mathbb{I}+|1\rangle\!\langle 1|\otimes w$,
where $\mathbb{I}$ is the identity operator and the tensor product
refers to the carrier subsystem (first component) and its phonon
environment (second component). As a result one gets
\begin{displaymath}
\widetilde{H} = W H W^{\dagger}
= -\tilde\Delta|1\rangle\!\langle 1|+H_{\mathrm{ph}},
\end{displaymath}
where 
$\tilde{\Delta}=\Delta+\sum_{\bm{k}}|g_{\bm{k}}|^{2}/(\hbar\omega_{\bm{k}})$
and 
$H_{\mathrm{ph}}
=\sum_{\bm{k}}\hbar\omega_{\bm{k}}b^{\dag}_{\bm{k}}b_{\bm{k}}$.

We assume that 
at the beginning ($t=0$) the state of the whole system is
$\rho_{0}=(|\Psi\rangle\!\langle \Psi |)\otimes \rho_{\mathrm{ph}}$,
where $\rho_{\mathrm{ph}}$ is the density matrix of the phonon
subsystem (environment) at thermal equilibrium and
$|\Psi\rangle$ is given by Eq.~(\ref{after-pulse}). For simplicity, we
will assume an equal superposition state, setting $\alpha=\pi/2$ and
$\varphi=-\pi/2$.

The evolution operator $U(t)=e^{-iHt/\hbar}$ may
be written as
\begin{eqnarray*}
U(t) & = & W^{\dagger}WU(t)W^{\dagger}W
= W^{\dagger}\widetilde{U}(t)W 
= W^{\dagger} \widetilde{U}(t) W\widetilde{U}^{\dagger} (t)
\widetilde{U}(t) \\
& = & W^{\dagger}  W(t)\widetilde{U}(t),
\end{eqnarray*}
where $ \widetilde{U}(t) =e^{-i\widetilde{H}t/\hbar}$ and 
$ W(t)=\widetilde{U}(t) W\widetilde{U}^{\dagger} (t)$.
Since $\widetilde{U}(t)$ is
 diagonal the explicit form of $W(t)$ may easily be found
and one gets
\begin{equation}
\label{op_phon}
U(t)=\left[ |0\rangle\!\langle 0|\otimes \mathbb{I}
+|1\rangle\!\langle 1|\otimes w^{\dag}w(t) \right]\widetilde{U}(t),
\end{equation}
where $w(t)=e^{-iH_{\mathrm{ph}}t/\hbar}we^{iH_{\mathrm{ph}}t/\hbar}$.

Using the evolution operator in the form (\ref{op_phon}) the system
state at a time $t$ may be written as
\begin{equation}
\label{mac_czas}
\rho(t)=\frac{1}{2}
\left(
\begin{array}{cc}
\rho_{\mathrm{E}}&
e^{-i\tilde{\Delta}t/\hbar} \rho_{\mathrm{E}}  w^{\dagger}(t) w\\
e^{i\tilde{\Delta}t/\hbar}w^{\dagger} w(t)\rho_{\mathrm{ph}} &
w^{\dagger} w(t)\rho_{\mathrm{ph}}  w^{\dagger}(t) w
\end{array}
\right),
\end{equation}
where we used the tensor product notation in which an
operator $A$ is expanded as $A=\sum_{m,n}|m\rangle\!\langle n|\otimes A_{mn}$ with
a set of operators $A_{mn}$ acting on the second subsystem, and
written in the matrix form with respect to the first subsystem.
The density matrix for the carrier subsystem is obtained by tracing
out the phonon degrees of freedom, i.e., 
$\rho_{\mathrm{S}}=\mathrm{Tr}_{\mathrm{E}}\rho$. Hence,
\begin{equation}
\label{dm-T}
\rho_{\mathrm{S}}(t)
=\frac{1}{2}\left(
\begin{array}{cc}
1& e^{-i\tilde{\Delta}t/\hbar}\langle  w^{\dagger}(t) w \rangle \\
e^{i\tilde{\Delta}t/\hbar}\langle w^{\dagger} w(t) \rangle&
1
\end{array}
\right).
\end{equation}
The average may be calculated 
using two rules for Weyl operators \cite{mahan00,roszak06b}. The
multiplication of two operators of the form 
\begin{equation}
\label{i12}
w_i=\exp\left[\sum_{\bm{k}}\left(g_{\bm{k}}^{(i)*}\gamma_{\bm{k}}^{\dagger}
-\gamma_{\bm{k}}^{(i)}b_{\bm{k}}\right)\right], \quad i=1,2,3
\end{equation}
yields
\begin{equation}
\label{multi}
w_1 w_2= w_3 \exp\left[\frac{1}{2}\sum_{\bm{k}}
\left(\gamma^{(1)*}_{\bm{k}} 
\gamma^{(2)}_{\bm{k}} -\gamma^{(1)}_{\bm{k}}\gamma^{(2)*}_{\bm{k}}\right)\right],
\end{equation}
where $w_{3}$ is given by Eq.~(\ref{i12}) with 
$\gamma^{(3)}_{\bm{k}}=\gamma^{(1)}_{\bm{k}}+\gamma^{(2)}_{\bm{k}}$.
The rule for averaging of an operator given by Eq.~(\ref{i12})
in the thermal equlibrium state is
\begin{equation}
\label{srednia_k}
\langle w_{i} \rangle=e^{-\frac{1}{2} \sum_{\bm{k}}
|\gamma_{\bm{k}}^{(i)} |^{2} (2n_{\bm{k}}+1)},
\end{equation}
where $n_{\bm{k}}$ are bosonic equilibrium occupation numbers.
The final result is
\begin{displaymath}
\langle  w^{\dagger}(t) w \rangle=
\exp\left\{-\sum_{\bm{k}} \left|\frac{g_{\bm{k}}}
{\hbar\omega_{\bm{k}}}\right|^{2}
\left[ i\sin\omega_{\bm{k}}  t 
+(1-\cos\omega_{\bm{k}}  t)(2n_{\bm{k}} +1)\right]\right\}.
\end{displaymath}

\begin{figure}[tb]
\centerline{\psfig{file=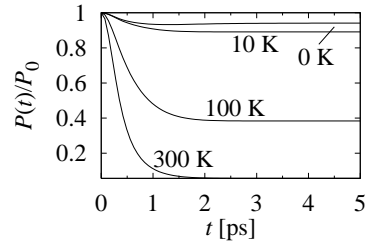,width=5cm}}
\caption{Decay of the coherent radiation from a confined exciton\index{exciton} at various
temperatures, as shown.
In our calculations we use typical parameters for a self-assembled
InAs/GaAs structure: single particle wave functions 
$\psi(\bm{r})$ modelled by
Gaussians with $l_{\bot}=4$~nm and $l_{z}=1$~nm [Eq.(\ref{ground-state})], 
the deformation
potential difference $\sigma_{\mathrm{e}}-\sigma_{\mathrm{h}}=9.5$ eV,
crystal density $\varrho=5300$ kg/m$^{3}$, and the speed of
sound $c=5150$ m/s.
} 
\label{fig:puredeph}
\end{figure}

The emitted coherent dipole
radiation is 
proportional to the non-diagonal element of the density matrix 
$\rho_s (t)$ and its amplitude is
\begin{equation}
\label{P}
P(t) = P_{0}|\langle  w^{\dagger}(t) w \rangle|.
\end{equation}
In Fig.~\ref{fig:puredeph} we show the
normalized polarization amplitude $P(t)/P_{0}$
(originally derived in Ref.~\refcite{krummheuer02}). The interaction
with the macroscopic crystal environment leads to a reduction of
coherent radiation due to pure dephasing of the exciton\index{exciton} state,
reflected by the reduced value of the non-diagonal element of the
density matrix $\rho_{\mathrm{S}}$. 
At $t=0$ one has $\langle  w^{\dagger}(t) w \rangle=1$, while at
large values of $t$, $\cos\omega_{\bm{k}}t$
oscillates very quickly as a function of $\bm{k}$ 
and averages to 0 (see also Ref.~\refcite{krummheuer02}). 
Thus, for long times, the polarization
amplitude tends to a temperature-dependent finite value 
\begin{displaymath}
P(t)\to P_{0}\exp\left[-\sum_{\bm{k}}
\left|\frac{g_{\bm{k}}}{\hbar\omega_{\bm{k}}}\right|^{2}
(2n_{\bm{k}}+1)\right]<P_{0}.
\end{displaymath}
This partial decay of coherence
is a characteristic feature of short-time
dephasing for carrier-phonon couplings encountered in real systems
\cite{borri01}.

\subsection{Perturbation theory for driven systems}
\label{sec:perturb}

In this section we derive the equations for the reduced density
matrix of the carrier subsystem to the leading order in the phonon
coupling, assuming that the unperturbed evolution is known. This
approach yields very simple and intuitive formulas that may easily be
applied to a range of problems. Here, we will again restrict the
formalism to a single exciton\index{exciton} level. A more general discussion is
given in Ref.~\refcite{grodecka05a}.

As already mentioned, the evolution
of the carrier subsystem is generated by the (time dependent)
Hamiltonian $H_{\mathrm{RWA}}$ [Eq.~(\ref{ham-2level-rwa})], 
describing the properties of the
system itself as well as its coupling to driving fields. 
The evolution of the phonon subsystem (reservoir) is
described by the Hamiltonian $H_{\mathrm{ph}}$ [the second term in 
Eq.~(\ref{ham-X-ph})].
The evolution operator for the
driven carrier subsystem and free phonon modes, without
the carrier-phonon interaction, is
\begin{displaymath}
    U_{0}(t)=U_{\mathrm{RWA}}(t)\otimes e^{-iH_{\mathrm{ph}}(t-s)},
\end{displaymath}
where $U_{\mathrm{RWA}}(t)$ is the operator for the unperturbed
evolution of the carrier subsystem and $s$ is the initial time.

The carrier-phonon coupling may be written in the form
$V=S\otimes R$,
where $S$ acts in the Hilbert space of the carrier
subsystem while the time-independent $R$ affect only the
environment. For instance, 
in the special case of Eq.~(\ref{ham-X-ph}), $S=|1\rangle\!\langle 1|$ and 
\begin{equation}\label{Rnn}
    R=\frac{1}{\sqrt{N}}\sum_{\bm{k}}
    g_{\bm{k}}\left(b_{\bm{k}}+b_{-\bm{k}}^{\dag}\right).
\end{equation}

We will assume that at the initial time $s$ the system is in the
product state
$\varrho(s)=|\psi_{0}\rangle\!\langle\psi_{0}|\otimes\rho_{\mathrm{ph}}$,
where $|\psi_{0}\rangle$ is a certain state of the carrier
subsystem and $\rho_{\mathrm{ph}}$ is the thermal equilibrium distribution
of phonon modes. Physically, such an assumption is usually
reasonable due to the existence of two distinct time scales: the
long one for the carrier decoherence (e.g. 1 ns ground state
exciton lifetime \cite{borri01,bayer02}) and the short one for the
reservoir relaxation (1 ps dressing time
\cite{borri01,krummheuer02,jacak03b}).

The starting point is the evolution equation for the density
matrix of the total system in the interaction picture with respect
to the externally driven evolution $U_{0}$, in the second order
Born approximation with respect to the carrier-phonon interaction
\cite{cohen98}
\begin{equation}\label{evol0}
    \tilde{\varrho}(t)=\tilde{\varrho}(s)
    +\frac{1}{i\hbar}\int_{s}^{t}d\tau[V(\tau),\varrho(s)]
    -\frac{1}{\hbar^{2}}\int_{s}^{t}d\tau\int_{s}^{\tau}d\tau'
      [V(\tau),[V(\tau'),\varrho(s)]],
\end{equation}
where
$\tilde{\varrho}(t)=U_{0}^{\dag}(t)\varrho(t)U_{0}(t),\;\;\;
    V(t)=U_{0}^{\dag}(t)VU_{0}(t)$
(it should be kept in mind that, in general, $V$ may depend on time itself).

The reduced density matrix of the carrier subsystem at time $t$ is
$\rho(t)=U_{\mathrm{RWA}}(t)\tilde{\rho}(t)U_{\mathrm{RWA}}^{\dag}(t),\;\;\;
    \tilde{\rho}(t)=\mathrm{Tr}_{\mathrm{R}}\tilde{\varrho}(t)$,
where the trace is taken over the phonon degrees of freedom.
The first (zeroth order) term in (\ref{evol0}) gives
\begin{equation}\label{ro0}
    \rho^{(0)}(t)
        =U_{\mathrm{RWA}}(t)|\psi_{0}\rangle\!\langle\psi_{0}|
            U_{\mathrm{RWA}}^{\dag}(t)
        =|\psi_{0}(t)\rangle\!\langle\psi_{0}(t)|.
\end{equation}
The second term vanishes, since it contains the thermal average of
an odd number of phonon operators. The third (second order) term describes
the leading phonon correction to the dynamics of the carrier
subsystem,
\begin{equation}\label{ro2}
    \tilde\rho^{(2)}(t)=
    -\frac{1}{\hbar^{2}}\int_{s}^{t}d\tau\int_{s}^{\tau}d\tau'
      \mathrm{Tr}_{\mathrm{R}}[V(\tau),[V(\tau'),\varrho(s)]].
\end{equation}

First of the four terms resulting from expanding the commutators
in (\ref{ro2}) is 
$(\mathrm{I})=-Q_{t}|\psi_{0}\rangle\!\langle\psi_{0}|$,
where
\begin{equation}\label{Q}
    Q_{t}=\frac{1}{\hbar^{2}}
    \int_{s}^{t}d\tau\int_{s}^{\tau}d\tau'
    S(\tau)S(\tau')
     \langle R(\tau-\tau')R\rangle.
\end{equation}
The operators $S$ and $R$ are transformed into the interaction
picture in the usual way
$ S(t)=U_{0}^{\dag}(t)SU_{0}(t)$,$R(t)=U_{0}^{\dag}(t)RU_{0}(t)$
and $\langle\hat{\mathcal{O}}\rangle
=\mathrm{Tr}_{\mathrm{R}}[\hat{\mathcal{O}}\rho_{\mathrm{ph}}]$ denotes the
thermal average (obviously $[U_{0}(t),\rho_{\mathrm{ph}}]=0$).

The second term is
\begin{displaymath}
    (\mathrm{II})= \frac{1}{\hbar^{2}}
    \int_{s}^{t}d\tau\int_{s}^{\tau}d\tau'
    |\psi_{0}\rangle\!\langle\psi_{0}|
    S(\tau')S(\tau)
    \langle R(\tau'-\tau)R\rangle
=-|\psi_{0}\rangle\!\langle\psi_{0}|Q^{\dag}_{t},
\end{displaymath}
where we used the relation   
$\langle R(\tau'-\tau)R\rangle=\langle R(\tau-\tau')R\rangle^{*}$.

In a similar manner, the two other terms may be combined into
\begin{displaymath}
    (\mathrm{III})+(\mathrm{IV})=
    \hat{\Phi}_{t}\left[|\psi_{0}\rangle\!\langle\psi_{0}|\right].
\end{displaymath}
where
\begin{equation}\label{Phi}
    \hat{\Phi}_{t}\left[\rho\right]=
    \frac{1}{\hbar^{2}}
    \int_{s}^{t}d\tau\int_{s}^{t}d\tau'
    S(\tau')\rho S(\tau)
     \langle R(\tau-\tau')R\rangle.
\end{equation}

In terms of the new Hermitian operators
\begin{equation}\label{Ah}
    A_{t}=Q_{t}+Q_{t}^{\dag},\;\;\;
    h_{t}=\frac{1}{2i}(Q_{t}-Q^{\dag}_{t}),
\end{equation}
the density matrix at the final time $t$ (\ref{ro0},\ref{ro2}) may
be written as
\begin{eqnarray}\label{master}
    \rho(t) & = & U_{\mathrm{RWA}}(t)\left(|\psi_{0}\rangle\!\langle\psi_{0}|
    -i\left[ h_{t},|\psi_{0}\rangle\!\langle\psi_{0}| \right]
\phantom{\frac{1}{2}}\right.\\
\nonumber
&& \left.
-\frac{1}{2}\left\{A_{t},|\psi_{0}\rangle\!\langle\psi_{0}|\right\}
    +\hat{\Phi}_{t}[|\psi_{0}\rangle\!\langle\psi_{0}|]\right)
    U^{\dag}_{\mathrm{RWA}}.
\end{eqnarray}
The first term is a Hamiltonian correction which does not lead to
irreversible effects and, in principle, may be compensated by
an appropriate modification of the control Hamiltonian
$H_{\mathrm{RWA}}$. The other two terms describe processes of
entangling the system with the reservoir, leading to the loss of
coherence of the carrier state.

Let us introduce the spectral density\index{spectral density} of the
reservoir, 
\begin{equation}\label{spdens}
    R(\omega)=\frac{1}{2\pi\hbar^{2}}\int dt
    \langle R(t)R\rangle e^{i\omega t}.
\end{equation}
For the operators given in Eq.~(\ref{Rnn}) one has explicitly
\begin{equation}\label{spdens-expli}
    R(\omega)=\frac{1}{\hbar^{2}}
    |n_{\mathrm{B}}(\omega)+1|\frac{1}{N}
    \sum_{\bm{k}}|g_{\bm{k}}|^{2}
    \left[\delta(\omega-\omega_{\bm{k}})
    +\delta(\omega+\omega_{\bm{k}}) \right],
\end{equation}
where $n_{\mathrm{B}}(w)=-n_{\mathrm{B}}(-\omega)-1$ is the Bose
distribution function. The spectral density $R(\omega)$ 
depends on the material
parameters and system geometry and characterizes the
properties of the lattice subsystem. For the deformation potential
coupling to LA phonons, it has the long-wavelength behavior 
$R(\omega)\sim\omega^{3}[n_{\mathrm{B}}(\omega)+1]$.

With the help of (\ref{spdens}) one may write
\begin{equation}\label{Fi}
    \hat{\Phi}_{t}\left[\rho\right]=    
    \int d\omega R(\omega)
    Y(\omega)\rho Y^{\dag}(\omega)
\end{equation}
where the frequency-dependent operators have been introduced,
\begin{equation}\label{Y}
    Y(\omega)
        =\int_{s}^{t}d\tau S(\tau)e^{i\omega \tau}.
\end{equation}
Using (\ref{spdens}) again one has
\begin{displaymath}
    Q_{t}=\int
    d\omega\int_{s}^{t}d\tau\int_{s}^{t}d\tau'
    \theta(\tau-\tau')S(\tau)S(\tau')
    R(\omega)e^{-i\omega(\tau-\tau')}.
\end{displaymath}
Next, representing the Heaviside function as
\begin{displaymath}
    \theta(t)=-e^{i\omega t}\int\frac{d\omega'}{2\pi i}
    \frac{e^{-i\omega' t}}{\omega'-\omega+i0^{+}},
\end{displaymath}
we write
\begin{eqnarray*}
    Q_{t} & = & -\int d\omega R(\omega)
    \int\frac{d\omega'}{2\pi i}
    \frac{Y^{\dag}(\omega')Y(\omega')}{\omega'-\omega+i0^{+}} \\
    & = & - \int d\omega R(\omega)
    \int\frac{d\omega'}{2\pi i} Y^{\dag}(\omega')Y(\omega')
    \left[-i\pi\delta(\omega'-\omega)+\mathcal{P}\frac{1}{\omega'-\omega}
    \right],
\end{eqnarray*}
where $\mathcal{P}$ denotes the principal value.

Hence, the two Hermitian operators defined in (\ref{Ah}) take the
form
\begin{equation}\label{A}
    A_{t}=
    \int d\omega
    R(\omega)Y^{\dag}(\omega)Y(\omega)
\end{equation}
and
\begin{equation}\label{h}
    h_{t}= \int d\omega R(\omega)
    \mathcal{P} \int\frac{d\omega'}{2\pi}
    \frac{Y^{\dag}(\omega')Y(\omega')}{\omega'-\omega}.
\end{equation}

In order to provide an example of an application of the perturbative
method presented in this section let us consider a $\pi/2$ rotation of
an excitonic two-level system in the presence of carrier-phonon coupling.
As we have already pointed out, any fast change of the state of
the carrier subsystem leads to spontaneous processes of lattice
relaxation that affect the coherence of the carrier state. It is
reasonable to expect that coherent control is recovered
if the evolution of the carrier subsystem is slow (adiabatic)
compared to the typical timescales of the lattice dynamics. Thus,
the requirement to avoid traces of carrier dynamics in the
outside world favors slow operation on the carrier subsystem,
contrary to other decoherence processes (of Markovian character),
like radiative decay of the exciton\index{exciton} or thermally activated
processes of phonon-assisted transitions to higher states. The
latter have the character of an exponential decay and, for short
times, contribute an error
$\tilde{\delta}=\tau_{\mathrm{g}}/\tau_{\mathrm{d}}$, where
$\tau_{g}$ is the gating time and $\tau_{d}$ is the decay time
constant. 
Here we consider the interplay between these two
contributions to the error for the solid-state qubit\index{qubit} 
implementation
using excitonic (charge) states in quantum dots (QDs)
\cite{biolatti00}, with computational states defined by the
absence ($|0\rangle$) or presence ($|1\rangle$) of one exciton\index{exciton} in
the ground state of the dot, operated by resonant coupling to
laser light. We show that it leads to a trade-off situation with a
specific gating time corresponding to the minimum decoherence for
a given operation\cite{alicki04a}.

To quantify the quality of the rotation, we use the \textit{fidelity}
\cite{nielsen00} 
\begin{equation}
\label{fidel}
F = \langle \psi_{0} |U_{\mathrm{\mathrm{RWA}}}^{\dag}(\infty)  
\rho (\infty) U_{\mathrm{RWA}}(\infty) |\psi_{0} \rangle^{1/2}
\end{equation}
which is a measure of the overlap between the ideal (pure) final 
state without perturbation, 
$U_{\mathrm{RWA}}(\infty)\psi_{0}$, and the actual final
state of the system given by the density matrix $\rho (\infty)$.
If the procedure is performed ideally, i.e. without discrepancies
from the desired qubit operation, then $F=1$.
The fidelity loss  $\delta = 1 - F^{2}$ is referred to 
as the \textit{error} of the quantum gate. 
Using the definition (\ref{fidel}) and the
Master equation (\ref{master}), the error may be written in a
general case as
\begin{displaymath}
    \delta=\left\langle \psi_{0}\left|A_{t}\right|\psi_{0}\right\rangle
      -\left\langle \psi_{0}\left|
            \hat{\Phi}\left[|\psi_{0}\rangle\!\langle\psi_{0}|
                \right]\right|
                \psi_{0}\right\rangle.
\end{displaymath}
It should be noted that the unitary correction generated by
$h_{t}$ does not contribute to the error at this order.
Using the definitions (\ref{Phi},\ref{A}) this may be further
transformed into
\begin{equation}\label{delta-detal}
    \delta=\int d\omega R(\omega)
    \left|\left\langle \psi_{\bot}|
    Y(\omega)|\psi_{0} \right\rangle\right|^{2},
\end{equation}
where $\psi_{\bot}$ is a state orthogonal
to $|\psi_{0}\rangle$ in the two-dimensional space of interest.

Since
the coherence of superpositions induced by short pulses is unstable
due to phonon-induced pure dephasing (Sec.~\ref{sec:ultrafast}),
it seems reasonable to perform operations on \textit{dressed
states}, i.e. on the correctly defined quasiparticles of the
interacting carrier-phonon system \cite{machnikowski07a}. This may be
formally achieved by employing the solid-state-theory concept of
adiabatic switching on/off the interaction\cite{pines89} (as done in
Ref.~\refcite{alicki02a}) to transform the states of
the noninteracting system into the states of the interacting one.
Thus, we assume adiabatic switching on/off of the interaction with
phonons by appending the appropriate exponent to the original
interaction Hamiltonian [Eq.~(\ref{ham-X-ph})],
\begin{eqnarray}
H_{\mathrm{int}}=e^{-\varepsilon |t|}\left[ |1\rangle\!\langle 1|
\sum_{\bm{k}}g_{\bm{k}} \left( b_{\bm{k}}+
b_{\bm{-k}}^{\dag} \right) \right], \label{int2}
\end{eqnarray}
where $\varepsilon=0^{+}$. The operator $S$ now becomes
\begin{displaymath}
    S(t)=U_{\mathrm{RWA}}(t)e^{-\varepsilon |t|}|1\rangle\!\langle
    1|U_{\mathrm{RWA}}^{\dag}(t),
\end{displaymath}
where the free evolution operator is generated by an optical pulse at
the resonance (Sec. \ref{sec:two-level-rabi}). The
general formula (\ref{delta-detal}) may now be used with the bare
initial state $|\psi_{0}\rangle$. The adiabatic procedure assures that it
is transformed to the dressed state before comparing it to the
density matrix $\rho$, so that the fidelity is defined with
respect to stable, dressed states.

The operator $Y(\omega)$ can be written in the form
\begin{eqnarray*}
Y(\omega) & = & \frac{1}{4i\omega}
 F(\omega)(|1\rangle\!\langle 0|-|0\rangle\!\langle 1|
    +|0\rangle\!\langle 0| - |1\rangle\!\langle 1|)  \\
 & & +\frac{1}{4i\omega}
 F^{*}(-\omega)(|0\rangle\!\langle 1|-|1\rangle\!\langle 0|
    +|0\rangle\!\langle 0| - |1\rangle\!\langle 1|),
\end{eqnarray*}
where
\begin{displaymath}
    F(\omega)=\int_{-\infty}^{\infty}d\tau e^{i\omega\tau}
        \frac{d}{d\tau}e^{i\Phi(\tau)}.
\end{displaymath}

Since in quantum information processing applications the
initial state of the quantum bit is in general not known, it is
reasonable to consider the error averaged over all input states.
Let us introduce the function
\begin{displaymath}
    S(\omega)=\omega^{2}|\langle\psi_{\bot}|
        Y(\omega)|\psi_{0}\rangle|^{2}_{\mathrm{av}},
\end{displaymath}
where the average is taken over the Bloch sphere.
According to Eq.~(\ref{delta-detal}), the
average error may now be written as
\begin{equation}\label{delta-av}
\delta = \int \frac{d \omega}{\omega^2}
R(\omega)S(\omega).
\end{equation}

The averaging is most conveniently performed by noting that
\begin{displaymath}
Y(\omega)=\frac{1}{2i\omega}F(\omega)|+\rangle\!\langle -|
+\frac{1}{2i\omega}F^{*}(-\omega)|-\rangle\!\langle +|,
\end{displaymath}
where $|\pm\rangle=(|0\rangle\pm |1\rangle)/\sqrt{2}$. Choosing
$|\psi_{0}\rangle = \cos\frac{\theta}{2}|+\rangle
  +e^{i\varphi}\sin\frac{\theta}{2}|-\rangle$, 
$|\psi^{\perp}_{0}\rangle = \sin\frac{\theta}{2}|+\rangle
  -e^{i\varphi}\cos\frac{\theta}{2}|-\rangle$,
one gets
\begin{displaymath}
S(\omega)=\frac{1}{4}\left|F(-\omega)e^{i\varphi}\cos^{2}\frac{\theta}{2}
-F^{*}(\omega)e^{-i\varphi}\sin^{2}\frac{\theta}{2}\right|^{2},
\end{displaymath}
which, upon averaging over the angles $\theta,\varphi$ on the
Bloch sphere, leads to
\begin{displaymath}
S(\omega)=\frac{1}{12}\left(|F(\omega)|^2+|F(-\omega)|^2\right).
\end{displaymath}

Let us now consider a Gaussian pulse for performing the quantum
gate,
$f(t)= [\alpha/(\sqrt{2\pi} \tau_{\mathrm{p}})]
    e^{-\frac{1}{2} ( {t/\tau_{\mathrm{p}} })^2}$,
were $\tau_{\mathrm{p}} $ is the gate duration, while $\alpha$ is
the angle determining the rotation. The
function $|F(\omega)|^2$ that carries all the needed information about
spectral properties of the system's dynamics may be approximately
written as
\begin{equation}\label{Fpm}
|F_{\pm}(\omega)|^{2}\approx\alpha^{2} e^{- \tau_{\mathrm{p}}^{2}
\left( \omega\pm \frac{\alpha}{\sqrt{2\pi}\tau_{\mathrm{p}}}
\right)^{2}}.
\end{equation}

\begin{figure}[tb]
\centerline{\psfig{file=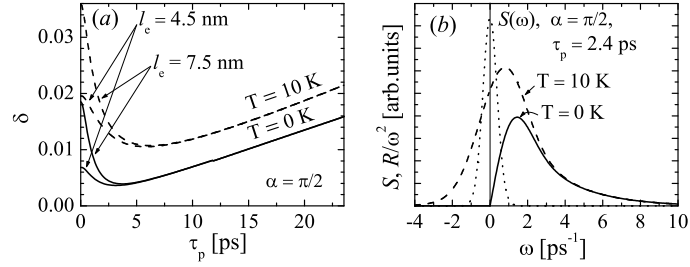,width=10cm}}
\caption{(a) Combined Markovian and
non-Markovian error for a $\alpha=\pi/2$ rotation on a qubit
implemented as a confined exciton\index{exciton} in a InAs/GaAs quantum dot, for
$T=0$ (solid lines) and $T=10$ K (dashed lines), for two dot sizes
(dot height is 20\% of its diameter). The Markovian decoherence
times are inferred from experimental data
\protect\cite{borri01}. (b) Spectral density of the phonon
reservoir $R(\omega)$ at these two temperatures and the gate
profile $S(\omega)$ for $\alpha=\pi/2$.}
\label{fig:tradeoff}
\end{figure}

As may be seen from (\ref{delta-av}) and (\ref{Fpm}), for a
spectral density $R(\omega)\sim \omega^{n}$ the error scales with
the gate duration as $\tau_{\mathrm{p}}^{-n+1}$ and
$\tau_{\mathrm{p}}^{-n+2}$ at low and high temperatures,
respectively. Therefore, for $n>2$ (typical for various types
of phonon reservoirs) the error
grows for faster gates. Assuming the spectral density of the form
$R(\omega)=R_{\mathrm{DP}}\omega^3$ for low frequencies (as for the
deformation potential coupling at low temperatures), we obtain from
(\ref{delta-av}) and (\ref{Fpm})
\begin{displaymath}
    \delta =
    \frac{1}{12} \alpha^2 R_{\mathrm{DP}}
        \tau_{\mathrm{p}}^{-2},\;\;\mbox{at}\;T=0
\end{displaymath}
This leading order formula holds for $\delta \ll 1$. Also, if we
introduce the upper cut-off, the error will be finite even for an
infinitely fast gate (see Fig.~\ref{fig:tradeoff}); this is the
ultrafast limit discussed in Sec.~\ref{sec:ultrafast}.

Thus, the phonon-induced error indeed decreases for 
slow driving. This could result in obtaining arbitrarily low
error by choosing a suitably low gate speed. However, if the
system is also subject to other types of noise this becomes
impossible \cite{alicki04a}. Indeed, assuming an additional contribution growing
with rate $\gamma_{\mathrm{M}}$, the total error per gate is
\begin{equation}\label{error-simple}
\delta=\frac{\gamma_{\mathrm{nM}}}{\tau_{\mathrm{p}}^{2}}
+\gamma_{\mathrm{M}}\tau_{\mathrm{p}}, \;\;
\gamma_{\mathrm{nM}}=\frac{1}{12}\alpha^{2}R_{\mathrm{DP}},\;\;
\gamma_{\mathrm{M}}=\frac{1}{\tau_{\mathrm{r}}},
\end{equation}
where $\tau_{\mathrm{r}}$ is the characteristic time of Markovian
decoherence (recombination time in the excitonic case). As a
result, the overall error is unavoidable and optimization is
needed. The formulas (\ref{error-simple}) lead to the optimal
values of the form (for $T=0$)
\begin{equation}
\delta_{\mathrm{min}}
=\frac{3}{2} \left(
\frac{2\alpha^{2}R_{\mathrm{DP}}}{3\tau_{\mathrm{r}}^{2}}
\right)^{1/3},\;\; \mbox{for}\; \tau_{\mathrm{p}}
=\left( \frac{2}{3}\alpha^{2}R_{\mathrm{DP}}\tau_{\mathrm{r}}
\right)^{1/3}. \label{eq:minerror}
\end{equation}

For the specific material parameters of GaAs, the optimal gate
time and minimal decoherence resulting from Eqs.
(\ref{eq:minerror}) are
\begin{displaymath}
\tau_{\mathrm{p}}=\alpha^{2/3} 1.47\; \mathrm{ps},\;\;
\delta_{\mathrm{min}}=\alpha^{2/3} 0.0035.
\end{displaymath}
The exact solution within the proposed model, taking into account
the cut-off and anisotropy (flat shape) of the dot and allowing
finite temperatures, is shown in Fig.~\ref{fig:tradeoff}. The
size-dependent cut-off is reflected by a shift of the optimal
parameters for the two dot sizes: larger dots allow faster gates
and lead to lower error.

It should be noted that these optimal times are longer than the
limits imposed by level separation
\cite{biolatti00,chen01,piermarocchi02}. Thus, the non-Markovian
reservoir effects (dressing) seem to be the essential limitation
to the gate speed. On the other hand, in the above discussion we used
simple Gaussian pulses and a straightforward way of encoding the
qubit. The error resulting from the phonon dynamics may be reduced by
optimizing the shape of the control pulse \cite{hohenester04,axt05a}
or by encoding the qubit into a state of an array of QDs \cite{grodecka06}.

\subsection{Correlation expansion}

\index{correlation expansion}The correlation expansion technique is a
standard method used for the 
description of quantum kinetics of interacting carriers and phonons in
semiconductor systems of any dimensionality \cite{rossi02}. It has
been successfully applied to carrier-phonon kinetics in quantum dots
driven by an optical field, beyond the instantaneous excitation limit
described in Sec.~\ref{sec:ultrafast} and beyond the weak perturbation
case which allows a perturbative treatment (Sec.~\ref{sec:perturb})
\cite{forstner03,krugel05,krugel06}. Compared to higher-dimensional
systems, in quantum dots coherent and non-equilibrium phonons play a
larger role because of the localized polaron effect. Therefore, a
reliable description of the carrier phonon-kinetics in these systems
requires a high enough degree of the correlation-expansion technique
\cite{krugel06}. 

Various implementations of this technique differ in notation and in
the choice of dynamical variables. Here, let us start from the three
dynamical variables $x,y,z$ describing the carrier state, 
$x=\langle\sigma_{x}(t)\rangle,\ldots$, where 
$\sigma_{i}(t)=e^{i\tilde{H}t/\hbar}\sigma_{i}
e^{i\tilde{H}t/\hbar}$ are Pauli
operators, written in the $|0\rangle,|1\rangle$ basis,
in the Heisenberg picture. 
These three variables are the coordinates of the evolving Bloch
vector, uniquely determining the reduced density matrix of the carrier
subsystem according to Eq.~(\ref{blochrep}).

From the Heisenberg equations of motion one
finds the dynamical equations for these three variables,
\begin{equation}
\label{x}
\dot{x}=i\langle[H,\sigma_{x}]\rangle
=-\Delta y-4y\sum_{\bm{k}}\mathrm{Re}\, 
B_{\bm{k}}-4y\sum_{\bm{k}}\mathrm{Re}\, y_{\bm{k}},
\end{equation}
and analogous for $y$ and $z$ (from now on, the time dependence will
not be written explicitly). Obviously, this set of equations is not
closed, but involves new the
phonon variables $B_{\bm{k}}=g_{\bm{k}}\langle b_{\bm{k}}\rangle$, as well as 
phonon-assisted variables of the form
$y_{\bm{k}}=g_{\bm{k}}\langle\langle \sigma_{y}b_{\bm{k}}\rangle\rangle=
\langle \sigma_{y}b_{\bm{k}}\rangle
-\langle \sigma_{y}\rangle\langle b_{\bm{k}}\rangle$. The double angular brackets,
$\langle\langle\ldots \rangle\rangle$,  denote the
correlated part of a product of operators, obtained by
subtracting all possible factorizations of the product. 

Next, one writes down the equations of motion for the new variables
that appeared in the previous step, for instance,
\begin{eqnarray}\label{yk}
\dot{y}_{\bm{k}} & = & i\langle[H,y_{\bm{k}}]\rangle
=\Delta x_{\bm{k}}-2Vz_{\bm{k}}-i\omega_{\bm{k}} y_{\bm{k}}
+|g_{\bm{k}}|^{2}(iyz+x)\\
\nonumber
&&+2\sum_{\bm{q}} (x_{\bm{q}\bm{k}}+\tilde{x}_{\bm{q}\bm{k}})
+4x_{\bm{k}}\sum_{\bm{q}}\mathrm{Re}\, B_{\bm{q}}
+2x\sum_{\bm{q}} (B_{\bm{q}\bm{k}}+\tilde{B}_{\bm{q}\bm{k}}),
\end{eqnarray}
where the new two-phonon and two-phonon-assisted variables are defined as
$B_{\bm{q}\bm{k}}=g_{\bm{q}}g_{\bm{k}}\langle\langle b_{\bm{q}}b_{\bm{k}}\rangle\rangle$, 
$\tilde{B}_{\bm{q}\bm{k}}=g_{\bm{q}}^{*}g_{\bm{k}}\langle\langle
b_{\bm{q}}^{\dag}b_{\bm{k}}\rangle\rangle$, 
$x_{\bm{q}\bm{k}}=g_{\bm{q}}g_{\bm{k}}\langle\langle \sigma_{x}b_{\bm{q}}b_{\bm{k}}\rangle\rangle$,
$\tilde{x}_{\bm{q}\bm{k}}=g_{\bm{q}}^{*}g_{\bm{k}}\langle\langle
\sigma_{x}b_{\bm{q}}^{\dag}b_{\bm{k}}\rangle\rangle$, etc. In the next step, one
writes the equation of motion for these new variables, introducing
three-phonon variables. It is clear that the resulting hierarchy of
equations in infinite and has to be truncated at a certain level. Here
we do this by setting all the correlated parts of the three-phonon and
three-phonon assisted variables equal to zero. This amounts to
neglecting the correlations involving three or more phonons or,
physically, to neglecting three-phonon processes (that is, emission or
absorption of three or more phonons within the memory time of the
phonon reservoir, which is of order of 1 ps).
The motivation for this
procedure is that higher order correlations should play a decreasing
role in the dynamics. From the equations of motion it is also clear
that such higher order correlations develop in higher orders with
respect to the coupling constants $g_{\bm{k}}$.

\begin{figure}[tb]
\centerline{\psfig{file=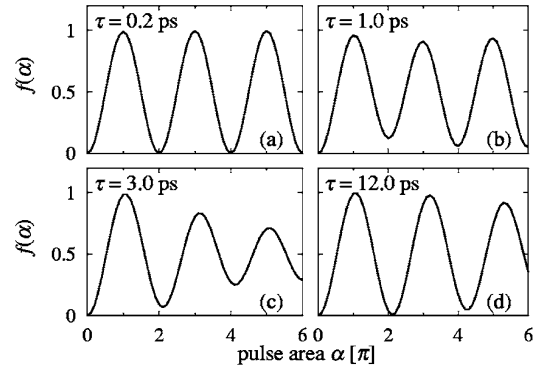,width=7cm}}
\caption{Rabi oscillations of the exciton\index{exciton}
occupation $f=(1+z)/2$ as a function of the nominal pulse area
$\alpha$, calculated using the correlation expansion at
$T=0$. Reprinted
from Ref.~\refcite{krugel06}, A. Krugel \textit{et al}.,
Phys. Rev. B. \textbf{73}, 035302 (2006), \copyright American
Physical Society 2006.}
\label{fig:rabi-ce}
\end{figure}

As an example of an application of this technique, 
Fig.~\ref{fig:rabi-ce} shows the results for the Rabi
oscillations of a coherently driven exciton\index{exciton} (see
Sec.~\ref{sec:two-level}) interacting with phonons \cite{krugel05}. An
interesting feature of this result is that the quality of oscillations
decreases for moderate pulse durations but then increases again for
longer pulses. This can be understood (using the perturbative
approach of Sec.~\ref{sec:perturb}) in terms of a resonance between
the oscillations of the charge distribution and the phonon modes
\cite{machnikowski04b}.  

\section{Conclusion}

This chapter presented an overview of some recent results
related to optical control and decoherence of carriers in
semiconductor quantum dots. The problems discussed here are of
interest not only from the scientific point of view but are also
important for possible applications of nanostructure-based devices in
the field of nano-electronics, optoelectronics and spintronics
\cite{scolnick04,bhattacharya04}.

Obviously, a chapter of limited length cannot give an exhaustive
review of this broad and rapidly developing field. Consequently, 
the goal of the
present review was rather to give an introduction to the field rather
than a complete, encyclopedic account of all the achievements. 
In particular, the two level (or few-level) model on which the
presented discussion was based, is obviously merely an
approximation to the complex semiconductor system. In fact, 
any interaction of the confined carriers with the
external driving fields not only leads to the desired quantum
transitions but also can induce some unwanted ones. If both the
desired and unwanted transitions have a discrete nature (e.g., the
exciton vs. biexciton\index{biexciton} transition in a QD) the latter may be
suppressed by a suitable choice of control pulses
\cite{chen01,piermarocchi02}. A more demanding problem of transitions
involving the macroscopic phonon continuum is treated in
Ref.~\cite{axt05a}. 

Another class of experiments and theories that are not covered by this
review concerns optical methods for coherent spin control in QDs.
The capability of encoding and manipulating information at the single-spin level
is of great importance for spintronic and quantum information
processing applications. Recent experimental progress includes
the generation \cite{dutt05} and optical control \cite{greilich06c,dutt06}
of the spin coherence together with a possible
read-out of the state of a single confined spin in a QD system \cite{atature07}.
It was also demonstrated that spin states in QDs may be prepared with high
fidelity (exceeding 99.8\%) 
by optical coupling of electronic spin states. This was done
by resonant excitation of the trion\index{trion} transition in the presence of
small heavy-light hole mixing \cite{atature06}.  
Corresponding to these achievements, a range of schemes
for optical spin control has been proposed theoretically
\cite{pazy03a,calarco03,troiani03,chen04,nazir04,lovett05}. 
These schemes are based on
quantum-optical procedures, exploiting the structure of selection
rules in a QD discussed in this chapter. 
In particular, the spin rotation is performed by coupling af a single
electron state to a trion\index{trion} state.
Since the spin control is
achieved by means of spin-dependent charge evolution these procedures
result in phonon-induced dephasing that may be described using the
methods described here \cite{roszak05b}. An introductory review of
optical spin control in QDs is given elsewhere \cite{machnikowski08a}.


\printindex                         
\end{document}